\documentclass[runningheads]{llncs}

\usepackage{graphicx}
\usepackage{cite}
\usepackage{amsmath,amssymb,amsfonts}
\usepackage{subfigure}
\usepackage{multirow}
\usepackage{wrapfig}

\begin{document}

\title{Efficient Quantification of Profile Matching Risk in Social Networks Using Belief Propagation}
\titlerunning{Efficient Quantification of Profile Matching Risk in Social Networks}
\author{Anisa Halimi \inst{1} \and Erman Ayday \inst{1,2}}
 \institute{Case Western Reserve University, Cleveland, OH, USA
 \and Bilkent University, Turkey \\
 \email{\{anisa.halimi,erman.ayday\}@case.edu}}
\maketitle             

\begin{abstract}
Many individuals share their opinions (e.g., on political issues) or sensitive information about them (e.g., health status) on the internet in an anonymous way to protect their privacy. However, anonymous data sharing has been becoming more challenging in today's interconnected digital world, especially for individuals that have both anonymous and identified online activities. The most prominent example of such data sharing platforms today are online social networks (OSNs). Many individuals have multiple profiles in different OSNs, including anonymous and identified ones (depending on the nature of the OSN). Here, the privacy threat is profile matching: if an attacker links anonymous profiles of individuals to their real identities, it can obtain privacy-sensitive information which may have serious consequences, such as discrimination or blackmailing. Therefore, it is very important to quantify and show to the OSN users the extent of this privacy risk. Existing attempts to model profile matching in OSNs are inadequate and computationally inefficient for real-time risk quantification.  Thus, in this work, we develop algorithms to efficiently model and quantify profile matching attacks in OSNs as a step towards real-time privacy risk quantification. For this, we model the profile matching problem using a graph and develop a belief propagation (BP)-based algorithm to solve this problem in a significantly more efficient and accurate way compared to the state-of-the-art. We evaluate the proposed framework on three real-life datasets (including data from four different social networks) and show how users' profiles in different OSNs can be matched efficiently and with high probability. We show that the proposed model generation has linear complexity in terms of number of user pairs, which is significantly more efficient than the state-of-the-art (which has cubic complexity). Furthermore, it provides comparable accuracy, precision, and recall compared to state-of-the-art. Thanks to the algorithms that are developed in this work, individuals will be more conscious when sharing data on online platforms. We anticipate that this work will also drive the technology so that new privacy-centered products can be offered by the OSNs. 
\keywords{Social networks \and profile matching \and deanonymization \and privacy risk quantification.}
\end{abstract}

\section{Introduction}

Many individuals, to preserve their privacy and to protect themselves against potential damaging consequences, choose to share content anonymously in the digital space. For instance, people share their opinions about different topics or sensitive information about themselves (e.g., their health status) without sharing their real identities, hoping that they will remain anonymous. Unfortunately, this is non-trivial in today's interconnected world, in which different activities of individuals can be linked to each other. An attacker, by linking anonymous activities of individuals to their real identities (via other publicly available and identified information about them), can obtain privacy-sensitive information about them. Thus, individuals need tools that show them the scale of their vulnerability against such privacy risks when they share content. In this work, we tackle this problem by focusing on data sharing on online social networks (OSNs). 

An OSN is a platform, in which, individuals share vast amount of information about themselves such as their social and professional life, hobbies, diseases, friends, and opinions. Via OSNs, people also get in touch with other people that share similar interests or that they already know in real-life~\cite{ellison2007social}. With the widespread availability of the Internet, OSNs have been a part of our lives more than ever. Most individuals have multiple OSN profiles for different purposes. Furthermore, each OSN offers different services via different frameworks, leading individuals share different types of information~\cite{ellison2007social}. Also, in some OSNs, users reveal their real identities (e.g., to find old friends), while in some OSNs, users prefer to remain anonymous (especially in OSNs in which users share anonymous opinions or sensitive information about themselves, such as their health status). Here, the privacy risk is the deanonymization of the anonymous OSN profile of a user using their other OSN profiles, in which the user is identified.

Such profile matching across OSNs (i.e., identifying profiles belonging to the same individuals) is a serious privacy threat, especially for individuals that have anonymous profiles in some OSNs and reveal their real identities in others. If an attacker can link anonymous profiles of individuals to their real identities, it can obtain privacy-sensitive information about individuals that is not intended to be linked to their real identities. Such sensitive information can then be used for discrimination or blackmailing. Thus, it is very important to quantify and show the risk of such profile matching attacks in an efficient and accurate way. 

An OSN can be characterized by (i) its graphical structure (i.e., connections between its users) and (ii) the attributes of its users (i.e., types of information that is shared by its users). The graphical structures of most popular OSNs show strong resemblance to social connections of individuals in real-life (e.g., Facebook). Existing work shows that this fact can be utilized to link accounts of individuals from different OSNs~\cite{narayan2}. However, without sufficient background information, just using graphical structure for profile matching becomes computationally infeasible. Furthermore, some OSNs or online platforms either do not have a graphical structure at all (e.g., forums) or their graphical structures do not resemble the real-life connections of the individuals (e.g., health-related OSNs such as PatientsLikeMe~\cite{patientslikeme}). In these types of OSNs, an attacker can utilize the attributes of the users for profile matching. Thus, to show the scale of the profile matching threat, it is crucial to process both the graphical structure and the other attributes of the users in an efficient and accurate way.

In this work, we efficiently model the profile matching problem in OSNs by considering both the graphical structure and other attributes of the users, a step towards delivering real-time information to OSN users about their privacy risks for profile matching due to their sharings on online platforms. 
Designing efficient privacy risk quantification tools is non-trivial, especially considering the scale of the problem. To overcome this challenge, we develop a novel, graph-based model generation algorithm to solve the profile matching problem in a significantly more efficient and accurate way than the state-of-the-art.

We formulate the profile matching problem as finding the marginal probability distributions of random variables representing the possible matches between user profile pairs from the joint probability distribution of many variables. We factorize the joint probability distribution into simpler local functions to compute the marginal probability distributions efficiently. To do so, we formulate the model generation for profile matching by using a graph-based algorithm. That is, we formulate the problem on a factor graph and develop a novel belief propagation (BP)-based algorithm to generate the model efficiently and accurately (compared to the state-of-the-art). The outcome of the model generation will pave the way towards developing real-time risk quantification tools (i.e., inform users about their privacy loss and its consequences as they share new content).

Our results show that the proposed model generation algorithm can match user profiles with an accuracy of up to $90\%$ (depending on the amount of information and attributes that users share). As more information is collected about the users profiles in social networks, the accuracy of the BP-based algorithm increases. Also, by analyzing the effect of social networks' size to obtained precision and recall values, we show the scalability of the proposed model generation algorithm.
We also show that by controlling the structure of the proposed graphical model, we can simultaneously improve the efficiency of the proposed model generation algorithm and increase its accuracy.

The rest of the paper is organized as follows. In Section~\ref{sec:related_work}, we discuss the related work. In Section~\ref{sec:threat}, we provide the threat model. In Section~\ref{sec:model}, we describe the proposed framework in detail. In Section~\ref{sec:evaluation}, we implement and evaluate the proposed framework using real-life datasets belonging to various OSNs. In Section~\ref{sec:discussion}, we discuss how the proposed scheme can be used for real-time privacy risk quantification, potential mitigation techniques, and generalization of the proposed scheme for different OSNs. Finally, in Section~\ref{sec:conclusion}, we conclude the paper. 
\vspace{-5pt}

\section{Related Work}\label{sec:related_work}

Several works in the literature have proposed profile matching schemes that leverage network structure, publicly available attributes of the users, or both of them. Profile matching based only on network (graph) structure is widely known as the deanonymization problem. Graph deanonymization (DA) attacks can be classified as (i) seed-based attacks~\cite{narayan2,nilizadeh,korula,ji2014structural,quantify}, in which a set of seeds (users' accounts in two different networks which belong to the same individual) are known; and (ii) seed-free attacks~\cite{wondracek,pedarsani}, in which no seeds are used. 
Narayanan and Shmatikov were among the first that proposed a graph deanonymization algorithm~\cite{narayan2}. Nilizadeh et al.~\cite{nilizadeh} improved the attack proposed by Narayanan et al. by proposing a community-level deanonymization attack. Korula and Lattanzi~\cite{korula} proposed a DA attack that by starting from a set of seeds, iteratively matches user pairs with the most number of neighboring mapped pairs. Ji et al.~\cite{quantify} quantified the deanonymizability of social networks from a theoretical perspective (i.e., focusing on social networks that follow a distribution model). 
Pedarsani et al.~\cite{pedarsani} proposed a Bayesian-based model to match users across social networks without using seeds. Their model uses node degrees and distances to other nodes. 
Sharad et al.~\cite{sharad} showed that users' re-identification (deanonymization) in anonymized social networks can be automated. 
Ji et al.~\cite{secgraph} evaluated several anonymization techniques and deanonymization attacks and showed that all state-of-art anonymization techniques are vulnerable to modern deanonymization attacks. 
Recently, Zhou et al.~\cite{zhou2018deeplink} proposed DeepLink, a deep neural network based algorithm that leverages network structure for user linkage.

Another line of works~\cite{iofciu,malhotra,nunes, vosecky,liu2013,jain,perito,motoyama,zafarani09,zafarani,goga13,liuarticle} have leveraged public information in the users' profiles (such as user name, profile photo, description, location, and number of friends) for profile matching. 
Shu et al.~\cite{shu} provided a broad review of the works that use public information for profile matching. 
Malhotra et al.~\cite{malhotra} built classifiers on various attributes to determine whether two user profiles are matching or not.
On the other hand, Zafarani et al.~\cite{zafarani} explored user name by analyzing the behaviour patterns of the users, the language used, and the writing style to link users across social media sites.
Goga et al.~\cite{goga13} showed that attributes that are hard to be controlled by users, such as location, activity, and writing style, may be sufficient for profile matching. 
Liu et al.~\cite{liu2014hydra} proposed a framework that mainly consists of three steps: behavior similarity modeling, structure consistency modeling, and multi-objective optimization. 
Goga et al.~\cite{goga2015reliability} conducted a detailed analysis of user profiles and their attributes identifying four properties: availability, consistency, non-impersonability, and discriminability. 
Andreou et al.~\cite{andreou} combined attribute and identity disclosure across social networks. 
Recently, Halimi et al.~\cite{halimi2017profile} proposed a more accurate profile matching framework based on machine learning techniques and optimization algorithms. 
One common thing about most of these aforementioned approaches is that they rely on training classifiers to determine whether a user pair is a match or not. 
We implemented some of these approaches and compared with the proposed framework in Section~\ref{sec:evaluation}. 

\noindent\textbf{Contribution of this paper:} Previous works show that there exists a non-negligible risk of matching user profiles on offline datasets. Showing the risk on offline datasets is not effective since users need tools that guide them at the time of data sharing in digital world. However, building algorithms that will pave the way towards real-time privacy risk quantification is non-trivial considering the scale of the problem. In this paper, we develop a novel belief propagation (BP)-based algorithm to generate the model efficiently and accurately (compared to the state-of-the-art). The proposed algorithm has linear complexity with respect to the number of user pairs (i.e., possible matches), while Hungarian algorithm~\cite{hungarian}, state-of-the-art that provides the highest accuracy (as shown in Section~\ref{sec:BP_results}), has cubic complexity with respect to the number of users. We also show that the proposed algorithm achieves comparable accuracy with the Hungarian algorithm while providing this efficiency advantage. 
\vspace{-5pt}
\section{Threat Model}\label{sec:threat}

We assume the attacker has access to user profiles in different OSNs. For simplicity we consider two OSNs: user profiles in OSN $A$ (the auxiliary OSN) are linked to their identities, while in OSN $T$ (the target OSN), the profiles of the individuals are anonymized. The attacker's goal is to match one or multiple user profiles from OSN $T$ to the profiles in OSN $A$ in order to determine the real identities of the users in OSN $T$. To do such profile matching, we assume that the attacker can only use the publicly available attributes of the users from OSNs $A$ and $T$. 

We study the extent of profile matching risk by means of two attacks: targeted attack and global attack. Targeted attack represents a scenario in which the attacker identifies the anonymous profile of a victim (or a set of victims) in OSN $T$ and aims to find the corresponding unanonymized profile of the same victim in OSN $A$. Global attack represents the case in which the attacker aims to link all profiles in OSN $T$ to their corresponding matches in OSN $A$.
\vspace{-5pt}
\section{Proposed Model Generation}\label{sec:model}

\begin{figure}[ht!]
\centering
\includegraphics[scale=0.4]{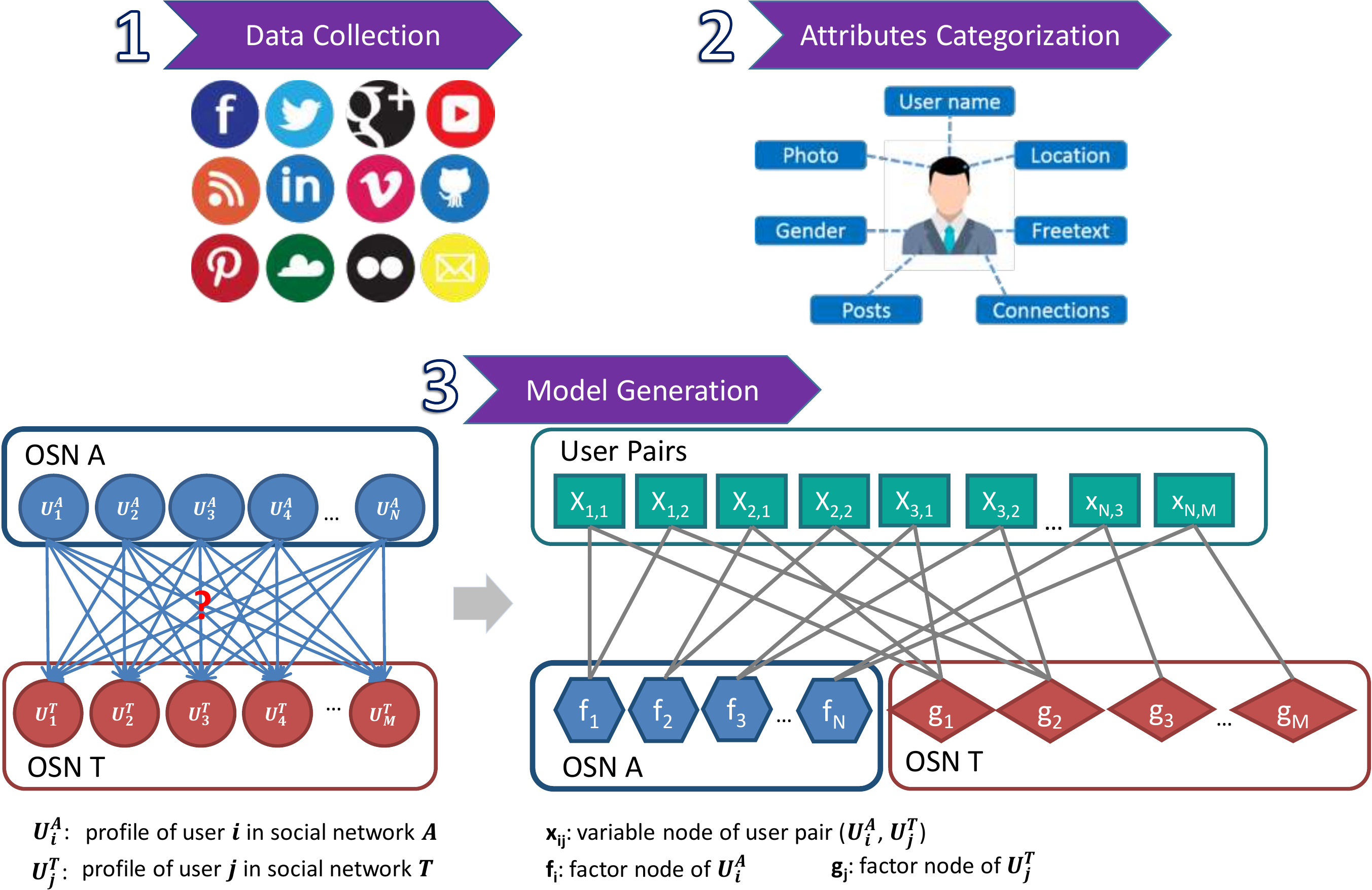}
\caption{Overview of the proposed framework. The proposed framework consists of 3 main steps: (1) data collection, (2) categorization of attributes and computation of attribute similarities, and (3) generation of the model.}
\vspace{-5pt}
\label{fig:workmodel}
\end{figure}

Let $A$ and $T$ represent the auxiliary and the target OSNs, respectively, in which people publicly share attributes such as date of birth, gender, and location. We represent the profile of a user $i$ in either $A$ or $T$ as $U_i^k$, where $k\in\{A,T\}$. We focus on the most common attributes that are shared in many OSNs and we categorize the profile of a user $i$ as $U_i^k=\{n_i^k,\ell_i^k,g_i^k,p_i^k,f_i^k,a_i^k,t_i^k,s_i^k,r_i^k\}$, where $n$ denotes the user name, $\ell$ denotes the location, $g$ denotes the gender, $p$ denotes the profile photo, $f$ denotes the freetext provided by the user in the profile description, $a$ denotes the activity patterns of the user (i.e., time instances at which the user posts), $t$ denotes the interests of the user (based on the sharings of the user), $s$ denotes the sentiment profile of the user, and $r$ denotes the (graph) connectivity pattern of the user. As discussed, the main goal of the attacker is to link the profiles between two OSNs. The overview of the proposed framework is shown in Figure~\ref{fig:workmodel}.
In the following, we describe the details of the proposed model generation algorithm.
\vspace{-5pt}

\subsection{Categorizing Attributes and Defining Similarity Metrics}~\label{sec:metrics}
Once the attributes of the users are extracted from their profiles, we first categorize them so that we can use them to compute the similarity values of attributes between different users. In this section, we define the similarity metrics for each attribute between a user $i$ in OSN $A$ and user $j$ in OSN $T$.

\noindent\textbf{User name similarity - $S(n^A_i,n^T_j)$:}
We use Levenshtein distance~\cite{levenshtein} to compute the similarity between user names of profiles. 

\noindent\textbf{Location similarity - $S(\ell^A_i,\ell^T_j)$:}
Location information collected from the users' profiles is usually text-based. We convert the textual information into coordinates via GoogleMaps API~\cite{googlemaps} and calculate the geographic distance between the corresponding coordinates.

\noindent\textbf{Gender similarity - $S(g^A_i,g^T_j)$:}
Availability of gender information is mostly problematic in OSNs. Some OSNs do not publicly share the gender information of their users. Furthermore, some OSNs do not even collect this information. In our model, if an OSN does not provide the gender information publicly (or does not have such information), we probabilistically infer the gender information by using a public name database. That is, we use the US social security name database\footnote{US social security name database includes year of birth, gender, and the corresponding name for babies born in the United States.} and look for a profile's name (or user name) to probabilistically infer the possible gender of the profile from the distribution of the corresponding name (among males and females) in the name database. We then use this probability as the $S(g^A_i,g^T_j)$ value between two profiles.

\noindent\textbf{Profile photo similarity - $S(p^A_i,p^T_j)$:}
We calculate the profile photo similarity through a framework named OpenFace~\cite{amos2016openface}. OpenFace is an open source tool performing face recognition. OpenFace first detects the face (in the photo), and then preprocesses it to create a normalized and fixed-size input for the neural network. The features that characterize a person's face are extracted by the neural network and then used in classifiers or clustering techniques. OpenFace notably offers higher accuracy than previous open source frameworks. Given two profile photos $p^A_i$ and $p^T_j$, OpenFace returns the photo similarity, $S(p^A_i,p^T_j)$, as a real value between $0$ (meaning exactly the same photo) and $4$.

\noindent\textbf{Freetext similarity - $S(f^A_i,f^T_j)$:}
Freetext data in an OSN profile could be a short biographical text or an ``about me'' page. We use NER (named-entity recognition)~\cite{ner} to extract features from the freetext information. The extracted features are location, person, organization, money, percent, date, and time. To calculate the freetext similarity between the profiles of two users, we use the cosine similarity between the extracted features from each user.

\noindent\textbf{Activity pattern similarity - $S(a^A_i,a^T_j)$:}
To compute the activity pattern similarity, we find the similarity between observed activity patterns of two profiles (e.g., likes or post). Let $a^A_i$ represent a vector including the times of last $|a^A_i|$ activities of user $i$ in OSN $A$. Similarly, $a^T_j$ is a vector including the times of last $|a^T_j|$ activities of user $j$ in OSN $T$. First, we compute the time difference between every entry in $a^A_i$ and $a^T_j$ and we determine $\min(|a^A_i|,|a^T_j|)$ pairs whose time difference is the smallest. Then, we compute the normalized distance between these $\min(|a^A_i|,|a^T_j|)$ pairs to compute the activity pattern similarity between two profiles.

\noindent\textbf{Interest similarity - $S(t^A_i,t^T_j)$:}
OSNs provide a platform in which users share their opinions via posts (e.g., tweets or tips), and this shared content is composed of different topics. In highlevel, first, we create a topic model using Latent Dirichlet Allocation (LDA)~\cite{Blei:2003:LDA:944919.944937}. Then, by using the created model, we compute the topic distribution of each post generated by the users of the auxiliary and the target OSNs. Finally, we compute the interest similarity from the distance of the computed topic distributions.

\noindent\textbf{Sentiment similarity - $S(s^A_i,s^T_j)$:}
Users typically express their emotions when sharing their opinions about certain issues on OSNs. To determine whether the shared text (e.g., post or tweet) expresses positive or negative sentiment we use sentiment analysis through Python NLTK (natural language toolkit) Text Classification~\cite{toolkit}. Given the text to analyze, the sentiment analysis tool returns the probability for positive and negative sentiment in the text. Users' moods are affected from different factors, so it is realistic to assume that they might change by time (e.g., daily). Thus, we compute the daily sentiment profile of each user, and daily sentiment similarity between the users. For this, first, we compute the normalized distribution of the positive and negative sentiments per day for each user, and then we find the normalized distance between these distributions for each user pair.

\noindent\textbf{Graph connectivity similarity - $S(r^A_i,r^T_j)$:}
As in~\cite{sharad}, for each user $i$, we define a feature vector $F_i=(c_0, c_1, ..., c_{n-1})$ of length $n$ made up of components of size $b$. Each component contains the number of neighbors that have a \textit{degree} in a particular range, e.g., $c_k$ is the count of neighbors with a degree in range [$k\cdot b, (k+1)\cdot b$]. We use the feature vector
length as 70 and bin size as 15 (as in~\cite{sharad}).
\vspace{-5pt}

\subsection{Generating the Model}\label{sec:hung}

We denote the set of profiles that are extracted for training from OSNs $A$ and $T$ as $\mathrm{A_t}$ and $\mathrm{T_t}$, respectively. Profiles are selected such that some profiles in $\mathrm{A_t}$ and $\mathrm{T_t}$ belong to the same individual. We let set $\mathrm{G}$ include pairs of profiles $(U_i^A,U_j^T)$ from $\mathrm{A_t}$ and $\mathrm{T_t}$ that belong to the same individual (i.e., coupled profiles). Similarly, we let set $\mathrm{I}$ include pairs of profiles that belong to different individuals (i.e., uncoupled profiles). For each pair of users in sets $\mathrm{G}$ and $\mathrm{I}$, we compute the attribute similarities based on the categorizations of the attributes (as discussed in Section~\ref{sec:metrics}). We label the pairs in sets $\mathrm{G}$ and $\mathrm{I}$ (as coupled and uncoupled) and add them to the training dataset. Then, to identify the weight (contribution) of each attribute, we use logistic regression.

Next, we select the profiles to be matched and construct the sets $\mathrm{A_e}$ (with size $N$) and $\mathrm{T_e}$ (with size $M$). Then, we compute the general similarity $S(U^{A}_i, U^{T}_j)$ between every user in $\mathrm{A_e}$ and $\mathrm{T_e}$ using the identified weights of the attributes to obtain the $N \times M$ similarity matrix $R$. Our goal is to obtain a one-to-one matching between the users in $\mathrm{A_e}$ and $\mathrm{T_e}$ that would also maximize the total similarity. One way of solving this problem is to formulate it as an optimization problem and use the Hungarian algorithm, a combinatorial optimization algorithm that solves the assignment problem in polynomial time~\cite{hungarian}. It is also possible to formulate profile matching as a classification problem and solve it using machine learning algorithms. Thus, we evaluate and compare the solution of this problem by using both the Hungarian algorithm and other off-the-shelf machine learning algorithms including k-nearest neighbor (KNN), decision tree, random forest, and SVM. 

Evaluations on different datasets (we will provide the details of the datasets later in Section~\ref{sec:datasetcreate}) show us that Hungarian algorithm provides significantly better precision, recall, and accuracy compared to other machine learning techniques (we will provide the details of our evaluation in Section~\ref{sec:BP_results}). However, assuming $N$ users in set $\mathrm{A_e}$ and $M$ users in set $\mathrm{T_e}$, the running time of the Hungarian algorithm for the above scenario is $O(\max\{N, M\}^3)$, and hence it is not scalable for large datasets. This raises the need for efficient, accurate, and scalable algorithms for model generation that will pave the way towards real-time privacy risk quantification.
\vspace{-5pt}

\subsection{Belief Propagation-Based Efficient Formulation of Model Generation}\label{sec:bp}

Inspired from the effective use of the message passing algorithms in information theory~\cite{pishro2005performance} and reputation management~\cite{ayday12}, in this research, for the first time, we formulate profile matching as an inference problem that infers the coupled profile pairs and develop an algorithm that relies on belief propagation (BP) on a graphical model. BP algorithm is based on a message-passing strategy for performing efficient inference using graphical models~\cite{pearl}. The problem we consider is different from~\cite{pishro2005performance,ayday12} and so is the formulation. In this section we formalize our approach and present the different components that are needed to quantify the profile matching risk. Our goal is to obtain comparable precision, recall, and accuracy values as in the Hungarian algorithm with significantly better efficiency. 

We represent the marginal probability distribution for a profile pair $(i,j)$ to be a coupled pair as $p(x_{i,j})$, where $x_{i,j}=1$ if profiles are matched as a result of the algorithm and $x_{i,j}=0$, otherwise. Then, we formulate the profile matching (i.e., determining if a profile pair is coupled or uncoupled) as computing the marginal probability distributions of the variables in set $\mathrm{X}=\{x_{i,j} : i \in A, j \in T\}$, given the similarity values between the user pairs in the similarity matrix $R$. Since the number of users in OSNs is high, it is computationally infeasible to compute the marginal probability distributions from the joint probability distribution $p(\mathrm{X}|R)$. Thus, we propose to factorize $p(\mathrm{X}|R)$ into local functions using a factor graph and run the BP algorithm to compute the marginal probability distributions in linear time (with respect to the number of profile pairs).

A factor graph is a bipartite graph containing two sets of nodes (variable and factor nodes) and edges between these two sets. We form a factor graph by setting a variable node for each variable $x_{i,j}$ (i.e., each profile pair). Thus, each variable node represents the marginal probability distribution of that profile pair being coupled or uncoupled. We use two types of factor nodes: (i) ``auxiliary'' factor node ($f_i$), representing each user $i$ in OSN $A$ and (ii) ``target'' factor node ($g_j$), representing each user $j$ in OSN $T$. Each factor node is connected to the variable nodes representing its potential matches. Factor nodes represent the statistical relationships between the user attributes and profile matching. Using the factor nodes, the joint probability distribution function can be factorized into products of several local functions, as follows:
\begin{equation}
p(X|R) = \frac{1}{Z}\left[\prod_{i=1}^N f_i(x_{\sigma f_i}, R) \prod_{j=1}^M g_j(x_{\sigma g_j}, R) \right],
\vspace{-3pt}
\end{equation}
where $\mathrm{Z}$ is a normalization constant, and $\sigma f_i$ (or $\sigma g_j$) represents the indices of the variable nodes that are connected to factor node $f_i$ (or $g_j$).

\begin{wrapfigure}{R}{0.46\textwidth}
	\centering
	\vspace{-7pt}
	\includegraphics[scale=0.33]{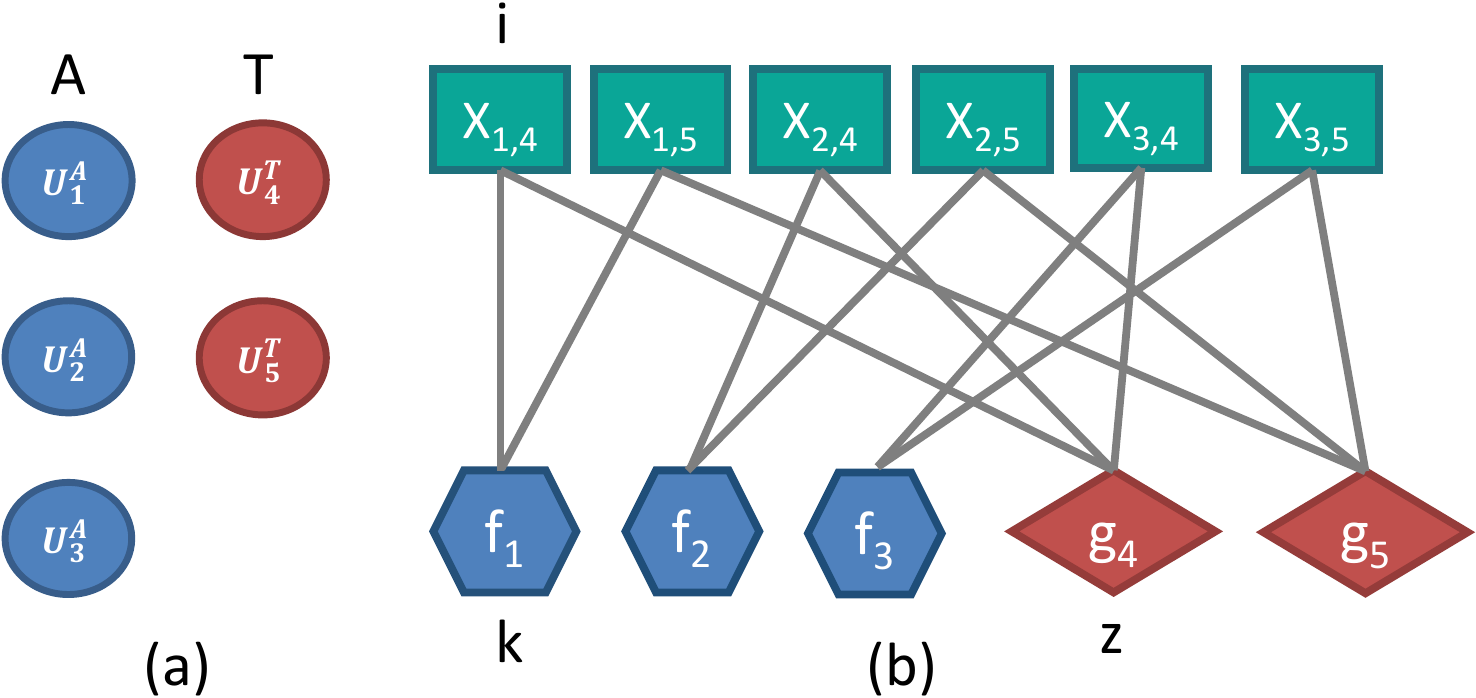}
	\caption{Factor graph representation of $3$ users from OSN $A$ and $2$ users from OSN $T$. (a) The users in both OSNs $A$ and $T$. (b) Factor graph representation of all the possible profile pairs combinations between users in OSNs $A$ and $T$.}
	\label{fig:bp_representation}
	\vspace{-5pt}
\end{wrapfigure}
Figure~\ref{fig:bp_representation} shows the factor graph representation of a toy example with 3 users from OSN $A$ and 2 users from OSN $T$. Here, each user corresponds to a factor node in the graph (shown as a hexagon or rhombus, respectively). Each profile pair is represented by a variable node and shown as a rectangle. Each factor node is connected to the variable nodes it acts on. For example $f_i$ is connected to all variable nodes (profile pairs) that contain $U_i^A$. The BP algorithm iteratively exchanges messages between the variable and the factor nodes, updating the beliefs on the values of the profile pairs (i.e., being a coupled or an uncoupled profile) at each iteration, until convergence.

Next, we introduce the messages between the variable and the factor nodes to compute the marginal distributions using BP. We denote the messages from the variable nodes to the factor nodes as $\mu$. We also denote the messages from the auxiliary factor nodes to the variable nodes as $\lambda$ and from the target factor nodes to the variable nodes as $\beta$. The message $\mu_{k \rightarrow i}^{(v+1)}\left(x_{i,j}^{(v)}\right)$ denotes the probability of $x_{i,j}^v=r$ ($r\in \{0, 1\}$), at the $v$th iteration. Also, $\lambda_{i \rightarrow k }^{(v)}\left(x_{i,j}^{(v)}\right)$ denotes the probability that $x_{i,j}^v=r$ $(r\in \{0, 1\}$) at the $v$th iteration given $\boldmath{R}$ (the messages $\beta$ can be also expressed similarly). In the following, we describe the message exchange between the variable node $x_{1,4}$, the auxiliary factor node $f_1$, and the target factor node $g_4$ in Figure~\ref{fig:bp_representation}. For clarity of presentation, we denote the variable and factor nodes $x_{1,4}$, $f_1$, and $g_{4}$ as $i$, $k$, $z$, respectively.

Following the general rules of BP~\cite{article:belief}, the variable node $i$ generates its message to auxiliary factor node $k$ by multiplying all the messages it receives from its neighbors, excluding $k$. Note that each variable node has only two neighbors (one auxiliary factor node and one target factor node). Thus, the message from the variable node $i$ to the auxiliary factor node $k$ at the $v$th iteration is as follows:
\begin{equation}
\mu_{i \rightarrow k}^{(v)}\left(x_{1,4}^{(v)}\right)=\beta_{z \rightarrow i}^{(v-1)}\left(x_{1,4}^{(v-1)}\right).
\vspace{-3pt}
\end{equation}
This computation is done at every variable node. The message from the variable node $i$ to the target factor node $z$ is also constructed similarly.

Next, factor nodes generate their messages. The message from the auxiliary factor node $k$ to the variable node $i$ is given by:
\begin{equation}
\lambda_{k \rightarrow i }^{(v)}\left(x_{1,4}^{(v)}\right)= \frac{1}{Z} \times S(U^A_1,U^T_4) \times \prod_{d \in(\sim i)} f_d\left(\mu_{d \rightarrow k}^{(v)} \left(x_{1,4}^{(v)}\right)\right),
\vspace{-3pt}
\end{equation}
where $(\sim i)$ means all variable node neighbors of $k$, except $i$. We compute function $f_d$ as:
\begin{equation}
f_d\left(\mu_{d \rightarrow k}^{(v)} \left(x_{1,4}^{(v)}\right)\right) = \left(1- \mu_{d \rightarrow k}^{(v)} \left(x_{1,4}^{(v)}\right)\right).
\vspace{-3pt}
\end{equation}
The above computation must be performed for every neighbor of each auxiliary factor node. The message from the target factor node $z$ to the variable node $i$ is also computed similarly. 

The next iteration is performed in the same way as the $v$th iteration. The algorithm starts at the variable nodes. In the first iteration (i.e., $v= 1$), all the variable nodes send to their neighboring factor nodes the same value ($\lambda_{i \rightarrow k }^{(1)}\left(x_{i,j}^{(1)}\right)=1/N$), where $N$ is the total number of ``auxiliary'' factor nodes.
The iterations stops when the probability distributions of all variables in $\mathrm{X}$ converge. The marginal probability distribution of each variable in $\mathrm{X}$ is computed by multiplying all the incoming messages at each variable node.
\vspace{-5pt}

\subsection{$\epsilon$-Accurate Model Generation}\label{sec:epsilon_accurate}

We also study the limitations and properties of the proposed BP-based model generation algorithm. We particularly analyze if the proposed algorithm maintains any optimality in any sense. For this, we use the following definition:
\begin{definition}{\textbf{$\epsilon$-accurate model generation.}}
We declare a model generation algorithm as $\epsilon$-accurate if it can match at least $\epsilon\%$ of the users accurately.
\end{definition}
Here, accuracy is the number of correctly matched coupled pairs by the proposed algorithm over the total number of coupled pairs. The above definition can also be made in terms of precision or recall (or both) of the proposed algorithm. Thus, for a fixed $\epsilon$, we study the conditions for an $\epsilon$-accurate algorithm. This also helps us understand the limits of profile matching in OSNs. To have an $\epsilon$-accurate algorithm with a high $\epsilon$ value, it can be shown that, we require the BP-based algorithm to iteratively increase the accuracy until it converges. This brings about the following sufficient condition about $\epsilon$-accuracy.
\begin{definition}{\textbf{Sufficient Condition.}}
Accuracy of the model generation algorithm increases with each successive iteration (until convergence) if for all coupled profiles $i$ and $j$, $Pr(x_{i,j}^{(2)}=1) > Pr(x_{i,j}^{(1)}=1)$ is satisfied.
\end{definition}
Depending on the fraction of the coupled profile pairs that meet the sufficient condition, $\epsilon$-accuracy of the proposed algorithm can be obtained. In Section~\ref{sec:evaluation}, we experimentally explore the cases in which this sufficient condition is satisfied with high probability.
\vspace{-5pt}

\section{Evaluation of the Proposed Mechanism}\label{sec:evaluation}

In this section, we evaluate the proposed BP-based algorithm by using real data from four OSNs. We also study the impact of various parameters to the $\epsilon$-accuracy of the proposed algorithm.
\vspace{-5pt}

\subsection{Evaluation Metrics}

To evaluate the proposed model, we mainly consider the global attack, in which the goal of the attacker is to match all profiles in $\mathrm{A_e}$ to all profiles in $\mathrm{T_e}$. In other words, the goal is to deanonymize all anonymous users in the target OSN (who have accounts in the auxiliary OSN). For the evaluation metrics, we use precision, recall, and accuracy. Hungarian algorithm and the proposed BP-based algorithm provide a one-to-one match between all the users. However, we cannot expect that all anonymous users in the target OSN have profiles in the auxiliary OSN. Therefore, some of the provided matches are useless for us. Thus, we select a ``similarity threshold'' (``probability threshold'' for machine learning techniques) for evaluation. Each matching scheme returns $1$ (i.e., true match) if the similarity/probability of user pair is higher than the threshold, and 0 otherwise. So, we consider as true positives the pairs that are correctly matched by the algorithm and whose similarity/probability is greater than the threshold. We also compute accuracy as the number of correctly matched coupled pairs identified by the algorithm over the total number of coupled pairs. 
\vspace{-5pt}

\subsection{Data Collection}\label{sec:datasetcreate}

To evaluate our proposed framework, we use three datasets: (i) Dataset~1 ($\mathrm{D1}$): Google+ - Twitter~\cite{halimi2017profile}, (ii) Dataset~2 ($\mathrm{D2}$): Instagram - Twitter, and (iii) Dataset~3 ($\mathrm{D3}$): Flickr social graph~\cite{Zafarani+Liu:2009}. 
To collect the coupled profiles in $\mathrm{D1}$ and $\mathrm{D2}$, social links in Google+ profiles and about.me (a social network where users provide links to their OSN profiles) were used, respectively. In terms of dataset sizes, (i) $\mathrm{D1}$ consists of $8000$ users in each OSNs where $4000$ of them are coupled profiles; (ii) $\mathrm{D2}$ consists of more than $10000$ coupled profiles (and more content about the OSN users compared to $\mathrm{D1}$); and (iii) $\mathrm{D3}$ consists of $50000$ users. 
In $\mathrm{D1}$, we use Twitter as our auxiliary OSN ($A$) and Google+ as our target OSN ($T$); in $\mathrm{D2}$, we use Twitter as our auxiliary OSN ($A$) and Instagram as our target OSN ($T$); and in $\mathrm{D3}$, we generate the auxiliary and the target OSN graphs as in~\cite{narayanan2011,sharad} by using a vertex overlap of $1$ and an edge overlap of $0.9$. 

\subsection{Evaluation Settings}\label{sec:settings}

Since the model generation process is the same for all three datasets, in the rest of the paper, we hold the discussion over a target and auxiliary network. From each dataset, we select $3000$ profile pairs ($1500$ coupled and $1500$ uncoupled) for training. We also select $500$ users from the auxiliary OSN and $500$ users from the target OSN to construct sets $\mathrm{A_e}$ and $\mathrm{T_e}$, respectively. Note that none of these users are involved in the training set. Among these profiles, we have $500$ coupled pairs and $249500$ uncoupled pairs, and hence the goal is to make sure that these $500$ users are matched with high confidence in a global attack scenario. Note that we do not use cross-validation because we consider all the possible user combinations to test our model and it is time-consuming to compute all similarity metrics for all combinations. Considering all the combinations instead of randomly selecting some user pairs is a more realistic evaluation setting since one can never know which users pairs the attacker will have access to. In cases that there are missing attributes (that are not published by the users) in the dataset, we assign a value for the attribute similarity based on the distributions of the attribute similarity values between the coupled and uncoupled pairs.

\subsection{Evaluation of BP-Based Model Generation}\label{sec:BP_results}

\begin{figure}[ht!]
	\centering
	\begin{subfigure}[$\mathrm{D1}$]{\includegraphics[scale=0.21]{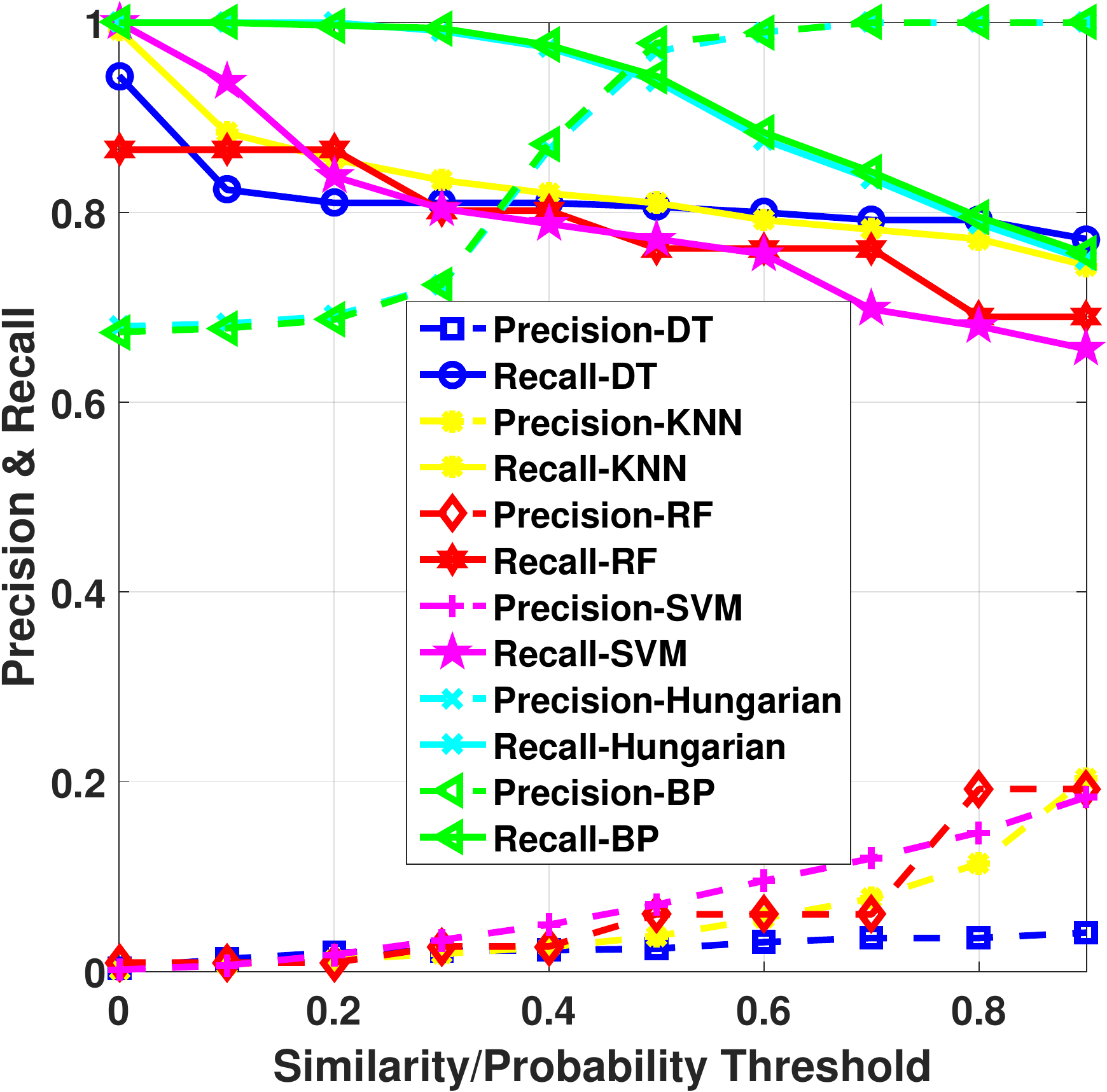}}
	\end{subfigure}\hfill
	\begin{subfigure}[$\mathrm{D2}$]{\includegraphics[scale=0.21]{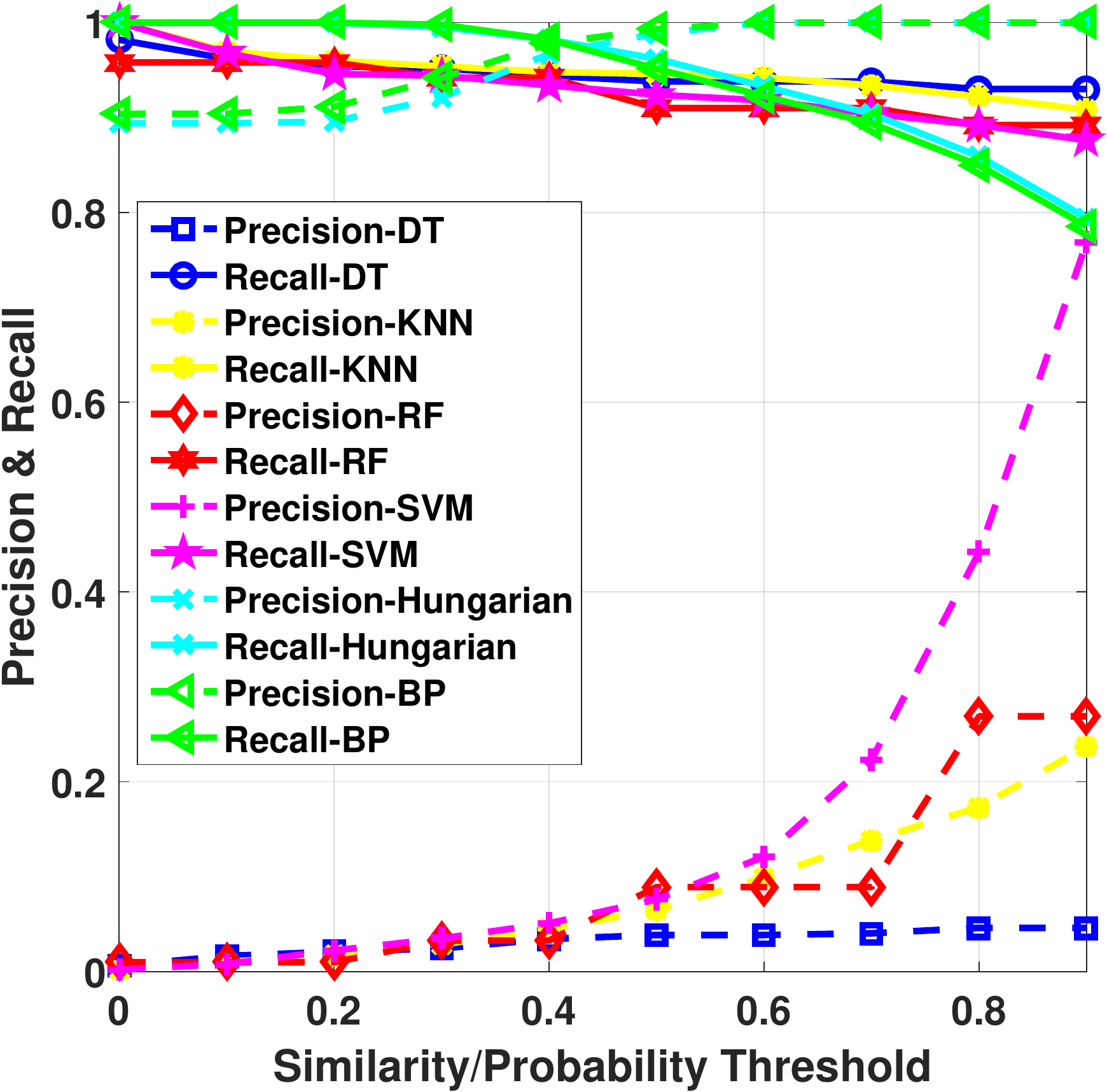}}
	\end{subfigure}\hfill
	\begin{subfigure}[$\mathrm{D3}$]{\includegraphics[scale=0.21]{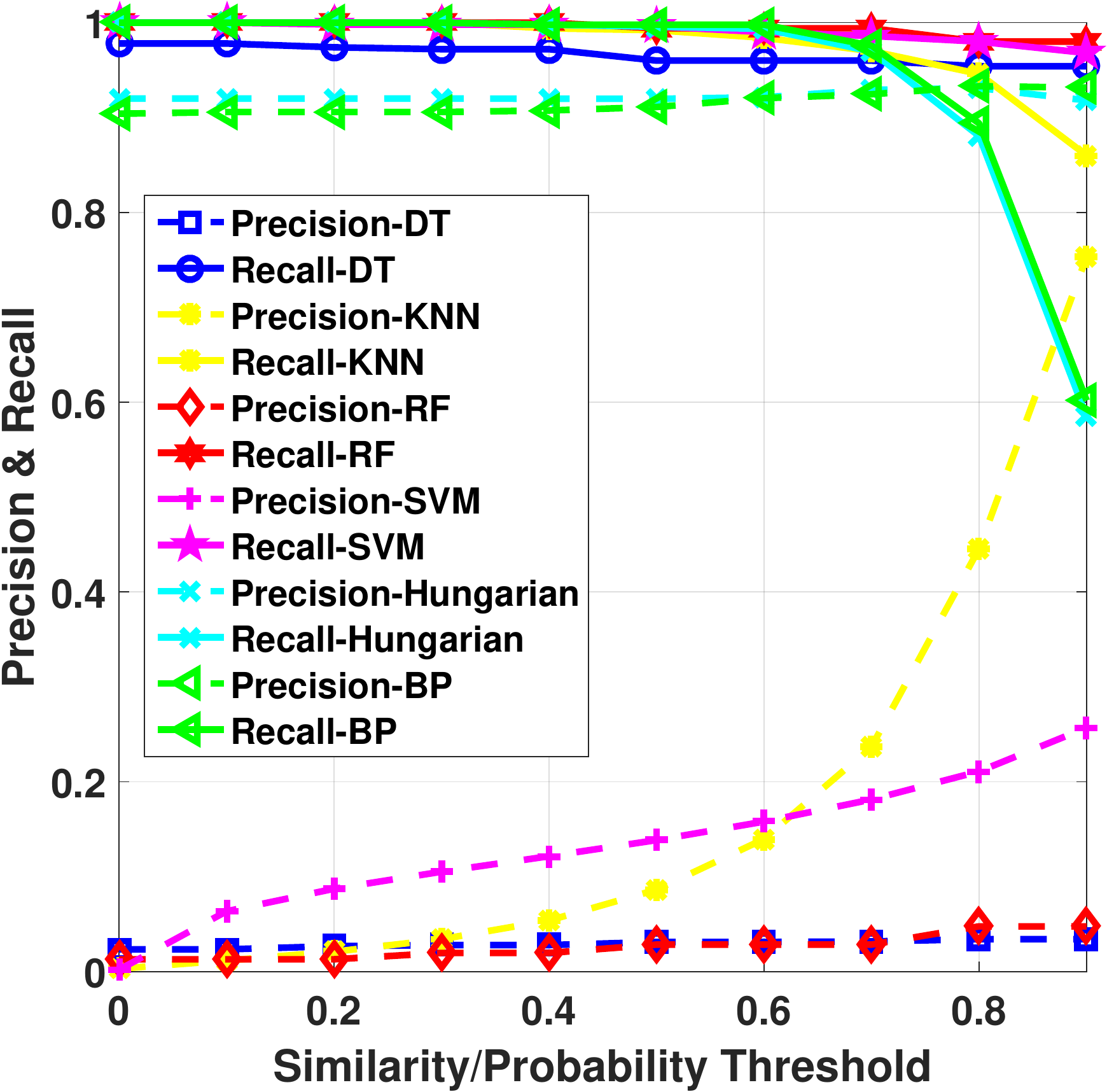}}
	\end{subfigure}\hfill
	\caption{Comparison of the proposed BP-based scheme with the Hungarian algorithm and machine learning techniques (decision tree - DT, KNN, random forest - RF, and SVM) in terms of precision and recall for $\mathrm{D1}$, $\mathrm{D2}$, and $\mathrm{D3}$.}
	\label{fig:BP_evaluation}
	\vspace{-10pt}
\end{figure}

In Figure~\ref{fig:BP_evaluation}, we show the comparison of the proposed BP-based model generation to~\cite{goga2015reliability,halimi2017profile,malhotra,nunes,sharad} for each dataset ($\mathrm{D1}$, $\mathrm{D2}$, and $\mathrm{D3}$). \cite{goga2015reliability,malhotra,nunes,sharad} use machine learning-based techniques (k-nearest neighbor (KNN), decision tree, random forest, and/or SVM), while~\cite{halimi2017profile} uses the Hungarian Algorithm. Our results show that the proposed scheme provides comparable precision and recall compared to the state-of-the-art Hungarian algorithm and it significantly outpowers machine leaning-based algorithms. For instance, the proposed algorithm provides a precision value of around $0.97$ (for a similarity threshold of $0.5$) in $\mathrm{D1}$. This means, if our proposed algorithm returns a matched profile pair that has a similarity value above $0.5$, the corresponding profiles belong to same individual with a high confidence. At the same time, the complexity of the proposed algorithm scales linearly with the number of user pairs, while the Hungarian algorithm suffers from cubic complexity. Note that precision and recall values obtained from $\mathrm{D2}$ are higher compared to the ones from $\mathrm{D1}$ as we collected more information about users in $\mathrm{D2}$. We also compare the BP-based model generation to the deep neural network based algorithm (DeepLink)~\cite{zhou2018deeplink} in $\mathrm{D3}$. For DeepLink, we use the same settings as in~\cite{zhou2018deeplink}. DeepLink achieves an accuracy of $84\%$ in $\mathrm{D3}$ which is slightly less than the one obtained by the proposed algorithm ($90\%$). DeepLink achieves a precision of $0.84$, and a recall of $1$ while the BP-based algorithm achieves a precision of $0.93$ and a recall of $0.9$. DeepLink provides a match for each user even if that user does not have a match.

\begin{figure}
        \centering
        \begin{subfigure}[$\mathrm{D1}$]{\includegraphics[scale=0.21]{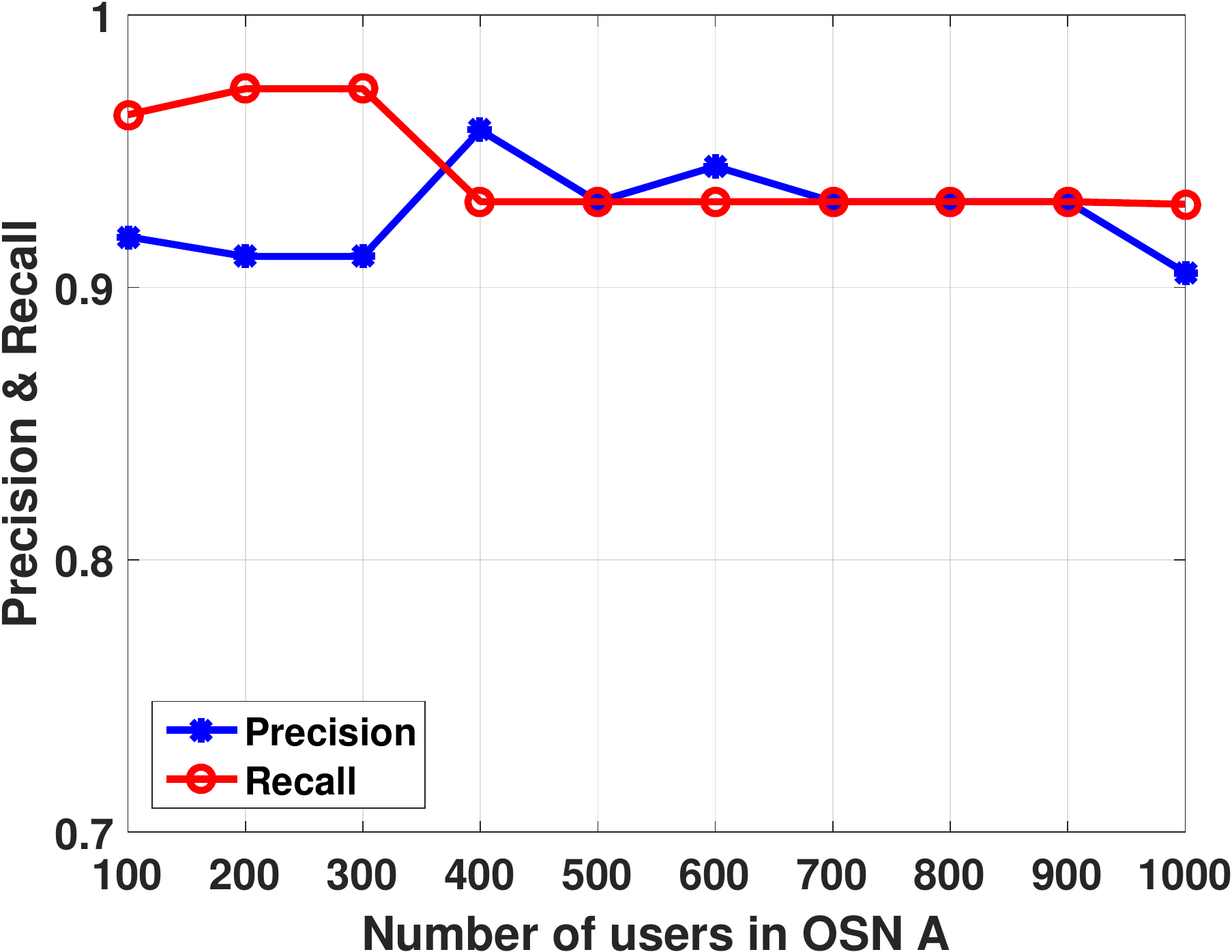}}
        \end{subfigure}\hfill
        \begin{subfigure}[$\mathrm{D2}$]{\includegraphics[scale=0.21]{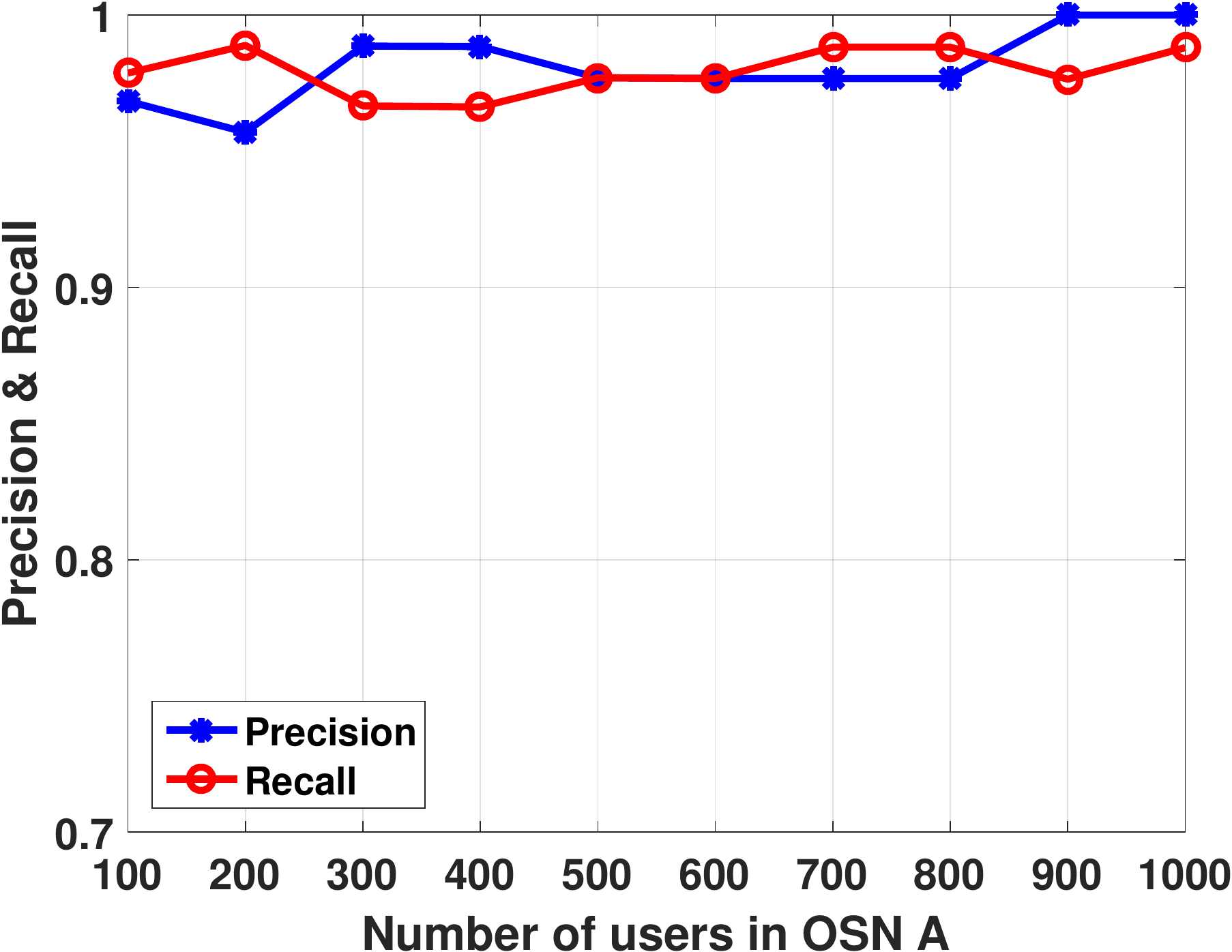}}
        \end{subfigure}\hfill
        \begin{subfigure}[$\mathrm{D3}$]{\includegraphics[scale=0.21]{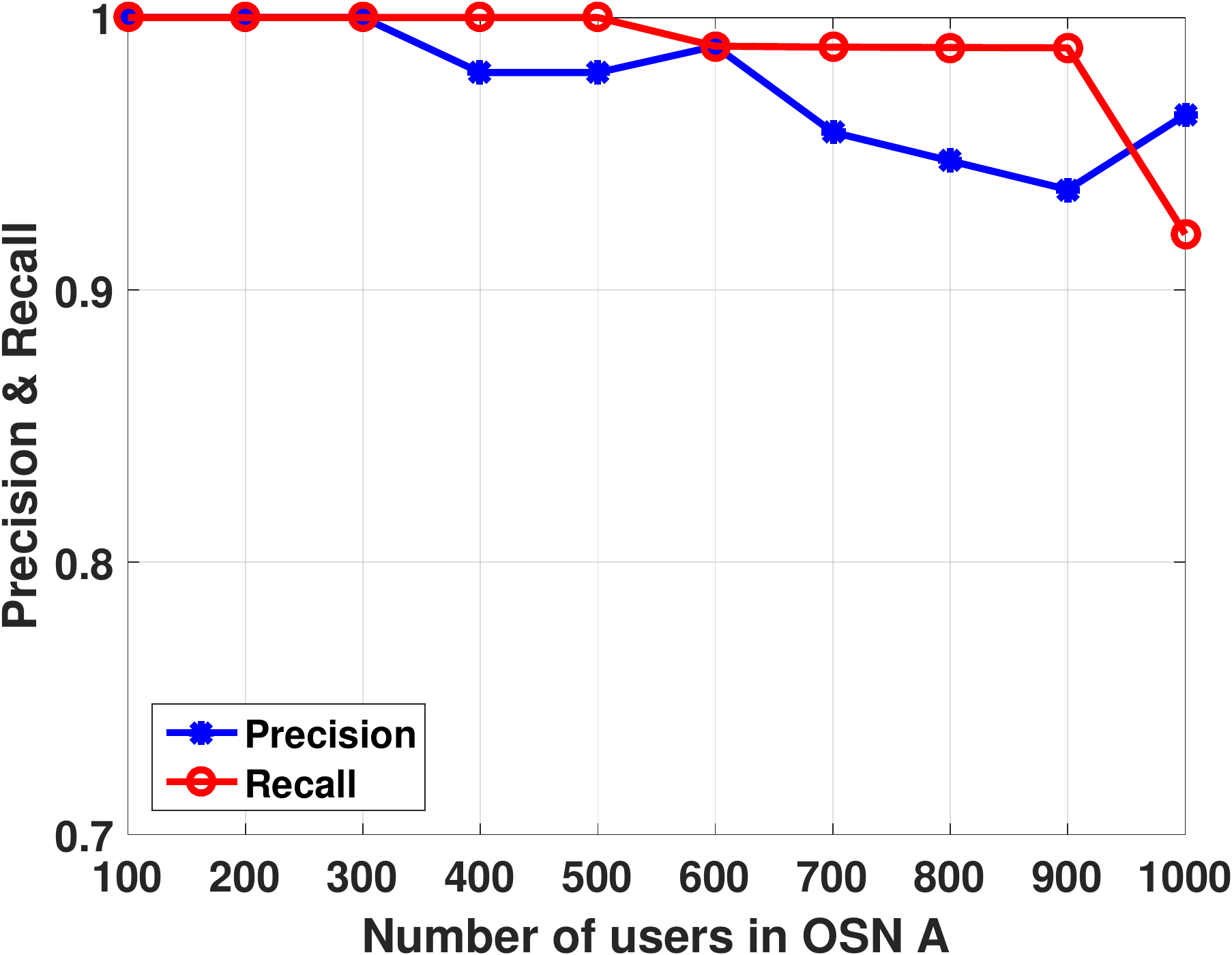}}
        \end{subfigure}\hfill
        \caption{The effect of auxiliary OSN's (OSN $A$) size to precision/recall when the size of target OSN (OSN $T$) is $100$ in $\mathrm{D1}$, $\mathrm{D2}$, and $\mathrm{D3}$.} 
        \label{fig:fixed_osn_t}
\vspace{-10pt}
\end{figure}

We also study the effect of the OSNs' size to precision and recall of the proposed algorithm. In Figure~\ref{fig:fixed_osn_t}, for each dataset, we show the precision/recall values of the BP-based algorithm when the number of users in the auxiliary OSN (OSN $A$) increases while the number of target users (i.e., users in OSN $T$) is fixed. We set the number of users in OSN $T$ as $100$ and increase the number of users in OSN $A$ from $100$ to $1000$ in steps of $100$. We observe that the precision/recall values of the proposed algorithm only slightly decrease with the increase of auxiliary OSN's size, which shows the scalability of our proposed algorithm. We achieve similar results for the other two scenarios: (i) when we fix the number of users in OSN $A$ and vary the number of users in OSN $T$; and (ii) when we increase the number of users in both OSNs $A$ and $T$.  Due to the space constraints, we present the details of the results of scenarios (i) and (ii) in Figures~\ref{fig:fixed_osn_a} and \ref{fig:both_osns}, respectively, in Appendix~\ref{app:scalability}.

Next, we evaluate the $\epsilon$-accuracy of the proposed model generation algorithm (introduced in Section~\ref{sec:epsilon_accurate}). There are many parameters to consider to analyze the $\epsilon$-accuracy of the proposed algorithm, such as the average degree of factor nodes, total similarity of each user in the target OSN with the ones in the auxiliary OSN, and number of users in the target and auxiliary OSNs. Here, we experimentally analyze and show the $\epsilon$-accuracy of the proposed algorithm considering such parameters. For evaluation, we use all datasets and pick $500$ users from OSN $T$. For all the studied parameters, we observe that at least $97\%$ of coupled profiles that can be correctly matched by the BP-based algorithm satisfy the sufficient condition (introduced in Section~\ref{sec:epsilon_accurate}). We observe that $\epsilon$ value is inversely proportional to the average degree of the factor nodes. In $\mathrm{D1}$, the $\epsilon$-accuracy of the proposed algorithm is $\epsilon=67$ and $\epsilon=84$ when the average degrees of the factor nodes are $500$ and $22$, respectively (we discuss more about the results of this experiment in Section~\ref{sec:complexity}).

\begin{wrapfigure}{R}{0.55\textwidth}
        \centering
        \begin{subfigure}[$\mathrm{D1}$]{\includegraphics[scale=0.17]{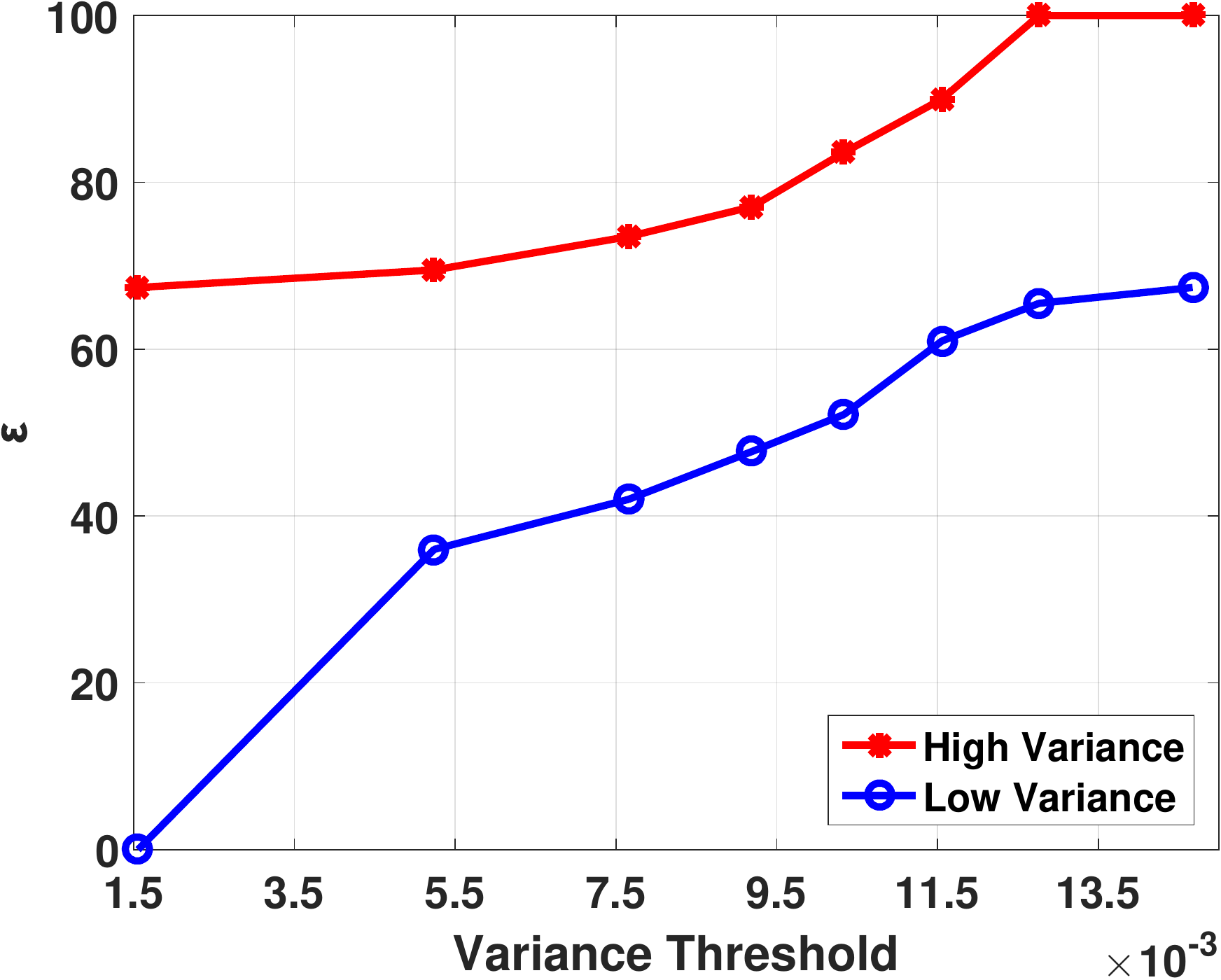}}
        \end{subfigure}\hfill
        \begin{subfigure}[$\mathrm{D2}$]{\includegraphics[scale=0.17]{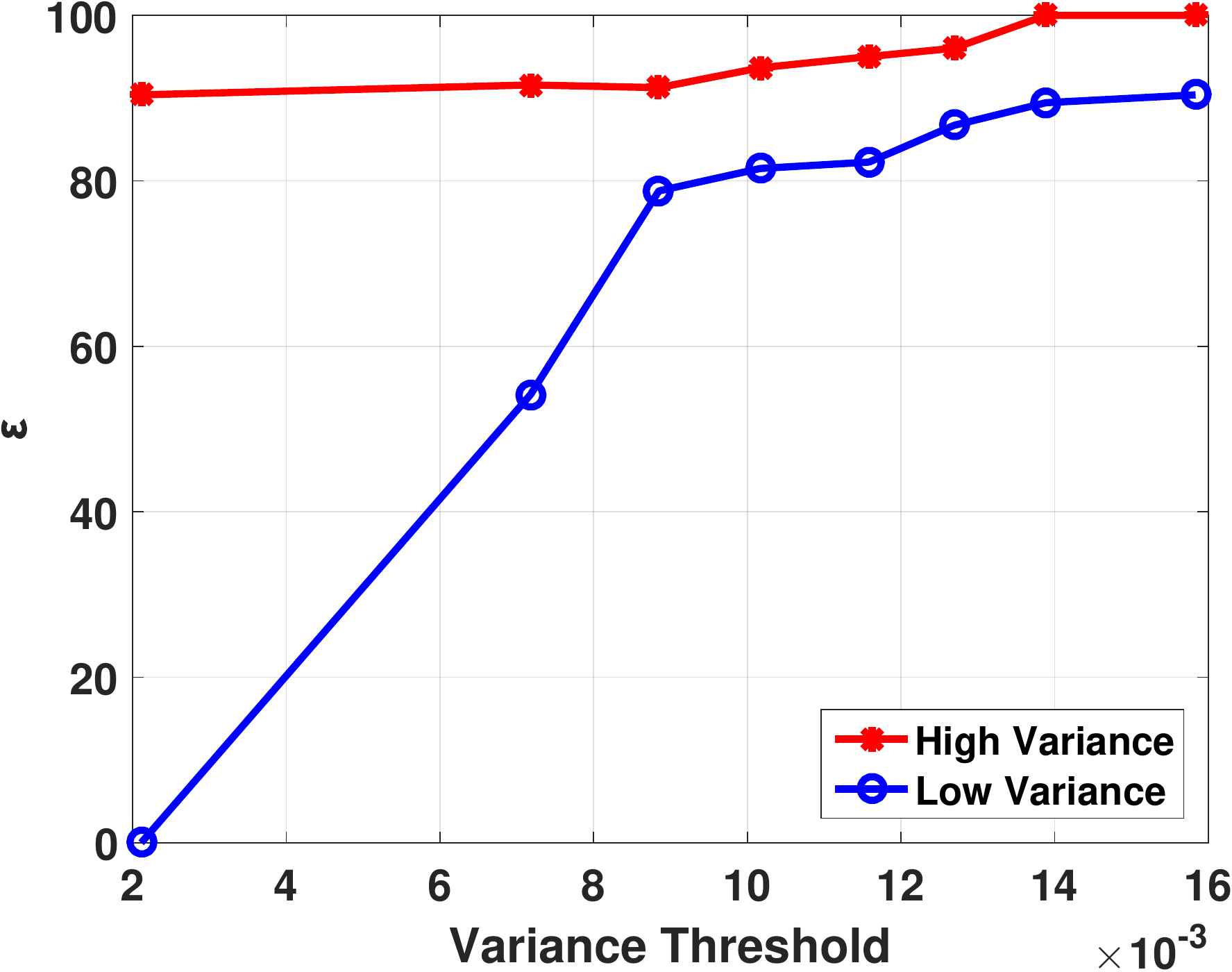}}
        \end{subfigure}\hfill
        \caption{The effect of variance threshold to $\epsilon$-accuracy in $\mathrm{D1}$ and $\mathrm{D2}$. For ``high variance'', our goal is to match users in OSN $T$ that have a variance (for the similarity values between a user in OSN $T$ and all users in OSN $A$) greater than the variance threshold, while for ``low variance'', we consider users that have a variance value smaller than the variance threshold.
        }
        \label{fig:var_sim}
    \vspace{-10pt}
\end{wrapfigure}
To study the impact of user pairs' similarity, for each user in OSN $T$, we compute the variance of the similarity values between that user and all users in OSN $A$. Then, we compute the accuracy of the proposed BP-based algorithm on users with varying variance values. For evaluation, we use $\mathrm{D1}$ and $\mathrm{D2}$. Our results show that $\epsilon$-accuracy of the proposed algorithm is higher for users with higher variances (as shown in Figure~\ref{fig:var_sim}). For instance, in $\mathrm{D1}$, we observe the $\epsilon$-accuracy as $\epsilon=42$ and $\epsilon=90$, when we run the proposed algorithm only for the users with low variance (lower than $0.008$) and high variance (higher than $0.012$), respectively. These results show that the vulnerability of the users in an OSN (for the profile matching attack) can be identified by analyzing particular characteristics of the OSNs. 

\begin{figure}[ht!]
        \centering
        \begin{subfigure}[$\mathrm{D1}$]{\includegraphics[scale=0.21]{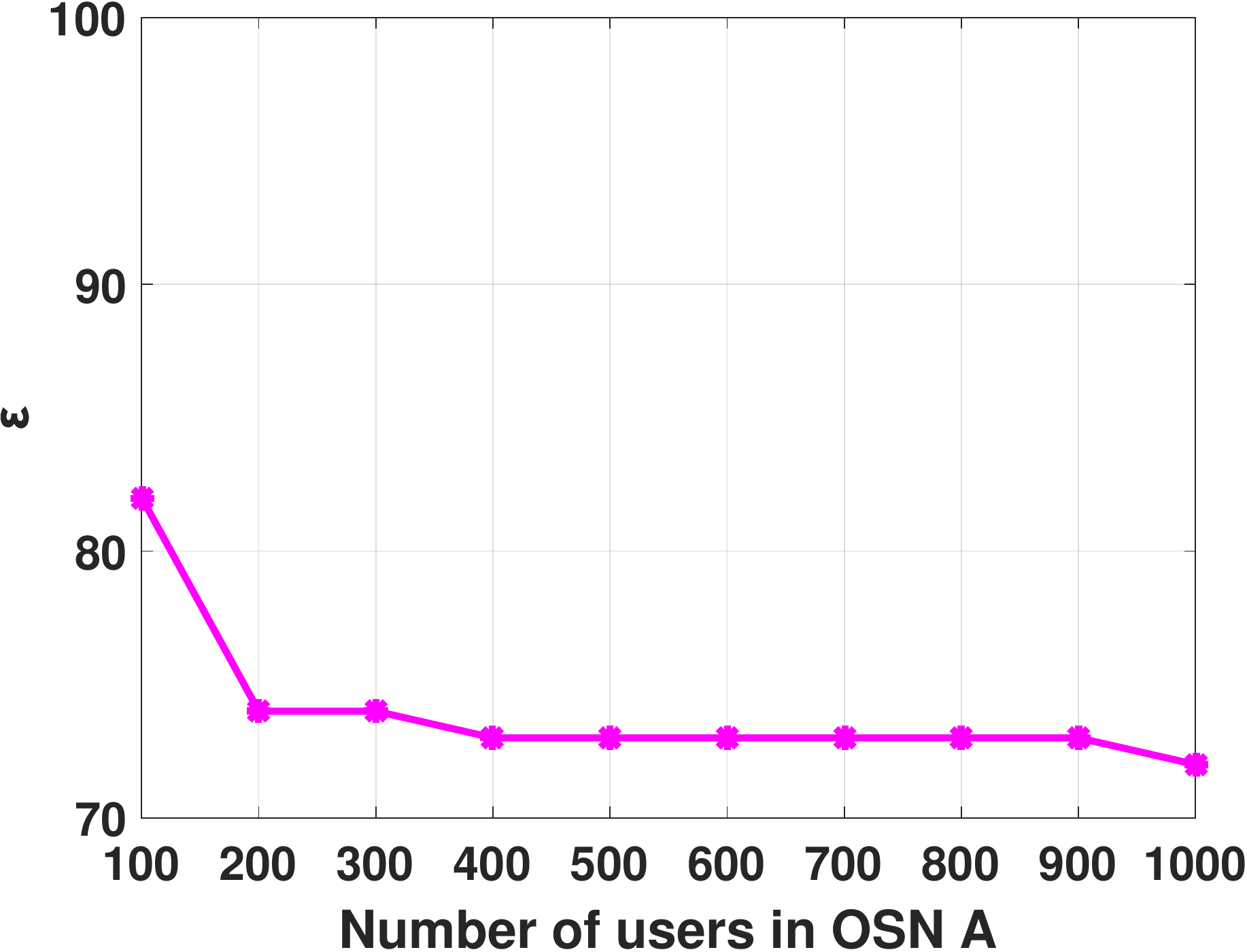}}
        \end{subfigure}\hfill
        \begin{subfigure}[$\mathrm{D2}$]{\includegraphics[scale=0.21]{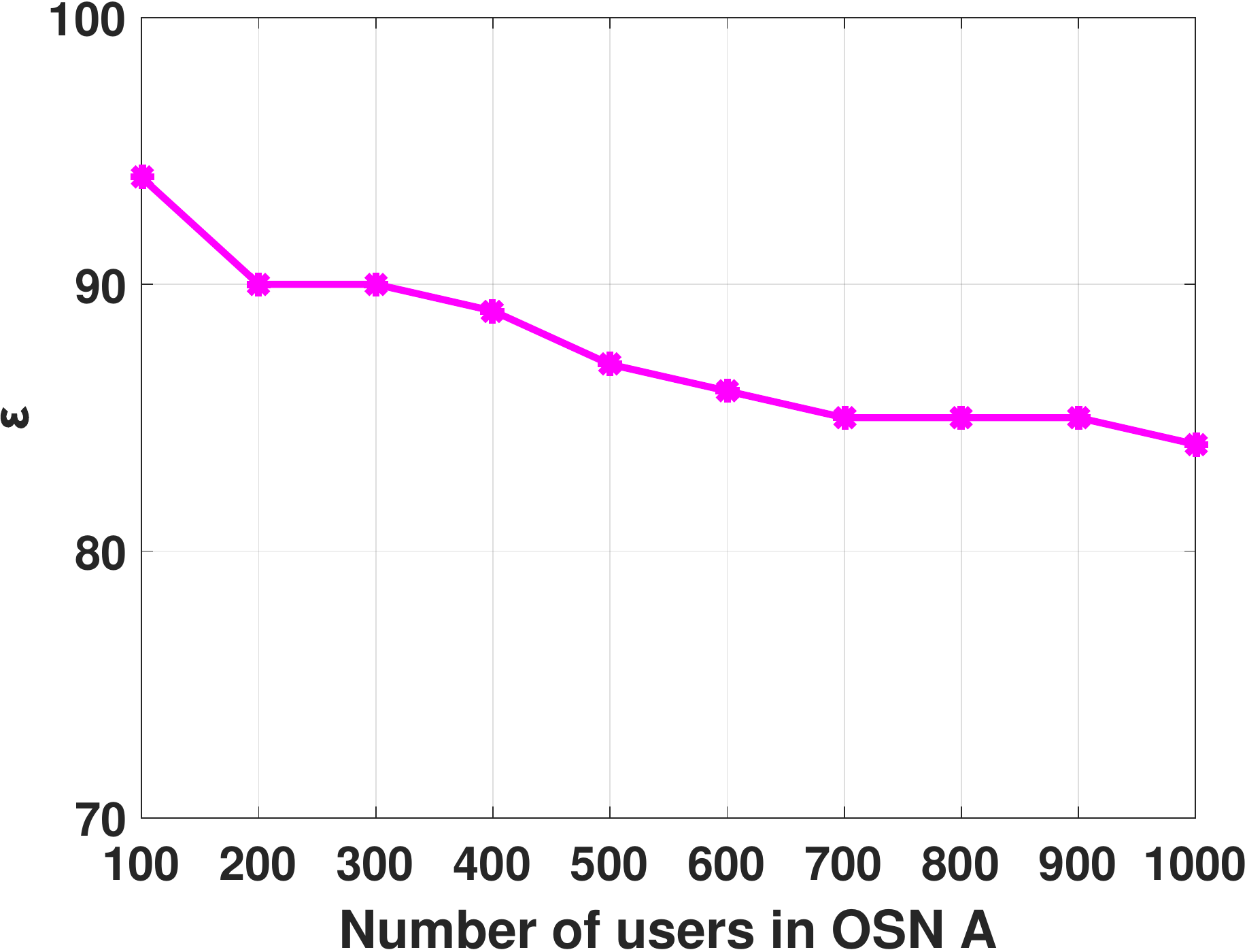}}
        \end{subfigure}\hfill
        \begin{subfigure}[$\mathrm{D3}$]{\includegraphics[scale=0.21]{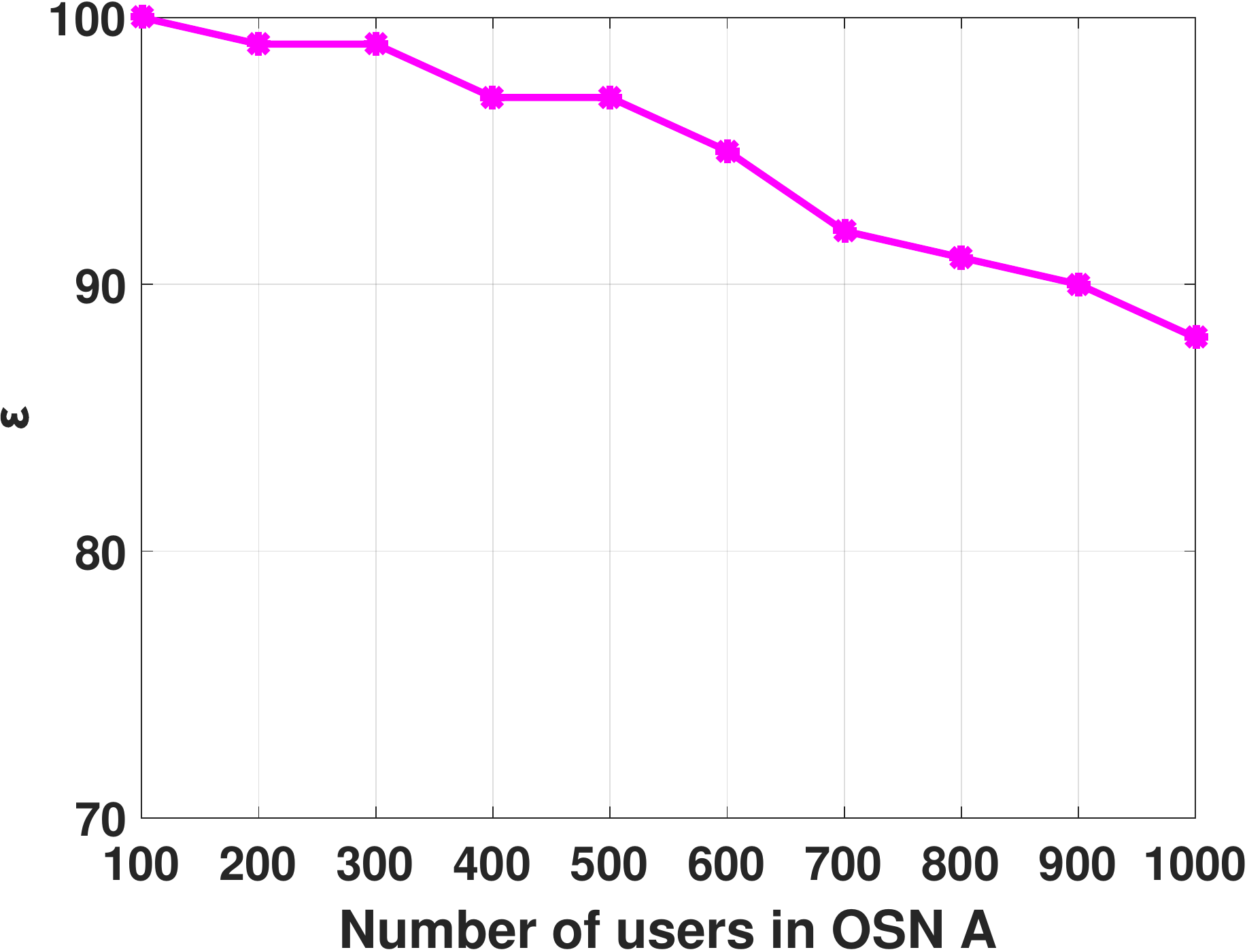}}
        \end{subfigure}\hfill
        \caption{The effect of auxiliary OSN's (OSN $A$) size to $\epsilon$-accuracy in $\mathrm{D1}$, $\mathrm{D2}$, and $\mathrm{D3}$. The size of target OSN (OSN $T$) is fixed to $100$.} 
        \label{fig:eps_accuracy}
\vspace{-10pt}
\end{figure}

Furthermore, we observe that $\epsilon$ value is inversely proportional to the number of users in OSN $A$ (as shown in Figure~\ref{fig:eps_accuracy}). The proposed algorithm achieves an $\epsilon$-accuracy of $\epsilon=82$ and $\epsilon=72$ in $\mathrm{D1}$ when the number of users in $A$ is $100$ and $1000$, respectively while the number of users in $T$ is $100$. This decrease in accuracy can be considered as low considering that the number of possible matches increases 10 times (from $10000$ to $100000$ user pairs). In $\mathrm{D2}$, we observe a similar trend to $\mathrm{D1}$. In $\mathrm{D3}$, accuracy decreases faster with the increase in number of users in OSN $A$ (compared to $\mathrm{D1}$ and $\mathrm{D2}$). This is because, in $\mathrm{D3}$, we only use the graph connectivity attribute for profile matching. Thus, as the number of users in OSN $A$ increases, the number of users with similar graph connectivity patterns also increases causing the decrease in accuracy.

\subsection{Complexity Analysis of the BP-Based Algorithm}\label{sec:complexity}

\begin{wraptable}{R}{0.55\textwidth}
	\centering 
	\resizebox{0.48\textwidth}{!}{
		\begin{tabular}{|c|c|c|c|c|}
            \hline
			 Dataset & Complexity & Precision & Recall & Accuracy \\ \hline
			 \multirow{3}{*}{$\mathrm{D1}$} & {$N^2$}  & $0.978$ & $0.944$ & $67.4\%$ \\ \cline{2-5}
			 & {$N\sqrt{N}$}  & $0.955$ & $0.939$ & $84.6\%$ \\ \cline{2-5}
			 &  {$N\log N$} & $0.975$ & $0.966$ & $96\%$  \\ \hline
			 \multirow{3}{*}{$\mathrm{D2}$} & {$N^2$}  & $0.978$ & $0.982$ & $90.4\%$ \\ \cline{2-5} 
			 & {$N\sqrt{N}$}  & $0.977$ & $0.954$ & $94.8\%$ \\ \cline{2-5}
			 &  {$N\log N$} & $0.934$ & $0.946$ & $90.6\%$  \\ \hline
			 \multirow{3}{*}{$\mathrm{D3}$} & {$N^2$}  & $0.925$ & $0.976$ & $90.4\%$ \\ \cline{2-5}
			 & {$N\sqrt{N}$}  & $0.92$ & $0.973$ & $91\%$ \\ \cline{2-5}
			 &  {$N\log N$} & $0.932$ & $0.976$ & $95\%$  \\ \hline
		\end{tabular}
	}
	\caption{Evaluation of the proposed BP-based algorithm with varying the number of variable nodes. $N$ denotes the number of users in OSN $T$, and each value in the complexity column shows the number of variable nodes (i.e., the number of users pairs used for profile matching).}
	\label{table:bp}
	\vspace{-10pt}
\end{wraptable}
The complexity of the BP-based algorithm is linear in the number of variable (or factor nodes). In the proposed BP-based algorithm, we generate a variable node for all potential matches between the target and the auxiliary OSNs. Assuming $N$ users in both target and auxiliary OSNs, this results in $N^2$ variable nodes in the graph. To analyze the effect of number of variable nodes on the performance, we experimentally try to change the graph structure and limit the number of variable nodes for each user in the target OSN. We heuristically decrease the average degree of the factor nodes from $N$ to $\sqrt{N}$ and $\log N$ by only leaving the variable nodes (user pairs) with the highest similarity values. For evaluation, we use all datasets ($\mathrm{D1}$, $\mathrm{D2}$, and $\mathrm{D3}$). We pick $500$ users from OSNs $A$ and $T$ to construct the test dataset (where there are $500$ coupled and $249500$ uncoupled profile pairs initially). In Table~\ref{table:bp}, we show the results with varying number of variable nodes. For instance, in $\mathrm{D1}$, we obtain an accuracy of $67.4\%$ when all the potential matches are considered (i.e., with $N$ variable nodes for each user in $T$) and an accuracy of $96\%$ when we use only $\log N$ variable nodes for each user. These results are important since they show that while reducing the complexity of the proposed BP-based algorithm, we can further improve its accuracy.
\vspace{-5pt}

\section{Discussion}\label{sec:discussion}

In this section we discuss how the proposed framework can be utilized for sensitive OSNs, and potential mitigation techniques against the identified profile matching risk. 

\subsection{Profile Matching on Sensitive OSNs}
Note that in $\mathrm{D1}$ and $\mathrm{D2}$, users provide the links to their social networks publicly. It is quite hard to obtain coupled profiles from social networks where users share sensitive information such as PatientsLikeMe. We expect to obtain similar results as long as users share similar attributes across OSNs. Considering that these users are more privacy-cautious, mostly non-obvious attributes such as interest, activity, sentiment similarity, or writing style can be used.

\subsection{Mitigation Techniques}

We foresee that the OSN can provide recommendations to the users (about the content they share) to reduce their risk for profile matching attacks. Such recommendation may include (i) generalizing or distorting some shared content of the user (e.g., generalizing the shared location or posting a content at a later time); or (ii) choosing not to share some content (especially for attributes that are hard to generalize or distort, such as interest or sentiment). When generating such recommendations, there are two main objectives: (i) content shared by the user should not increase user's risk for profile matching and (ii) utility of the content shared by the user (or utility of user's profile) should not decrease due to the applied countermeasures. Using a utility metric for the user's profile, the proposed framework (in Section~\ref{sec:bp}) can be used to formulate an optimization between the utility of the user's profile and privacy of the user. The solution of this optimization problem can provide recommendations to the user about how to (or whether to) share a new content on their profile. 

\section{Conclusion}\label{sec:conclusion}

In this work, we have proposed a novel message passing-based framework to model the profile matching risk in online social networks (OSNs). We have shown via simulations that the proposed framework provides comparable accuracy, precision, and recall compared to the state-of-the-art, while it is significantly more efficient in terms of its computational complexity. We have also shown that by controlling the structure of the proposed BP-algorithm we can further decrease the complexity of the algorithm while increasing its accuracy. We believe that the proposed framework will be instrumental for OSNs to educate their users about the consequences of their online sharings. It will also pave the way towards real-time privacy risk quantification in OSNs against profile matching attacks.  

\vspace{5pt}
\noindent\textbf{Acknowledgment.} We would like to thank the anonymous reviewers and our shepherd Shujun Li for their constructive feedback which has helped us to improve this paper.

\bibliographystyle{splncs04}
\bibliography{references}
\appendix
\section*{Appendix}
\section{Scalability of the BP-Based Algorithm}\label{app:scalability}

\begin{figure}[ht!]
        \centering
        \begin{subfigure}[$\mathrm{D1}$]{\includegraphics[scale=0.21]{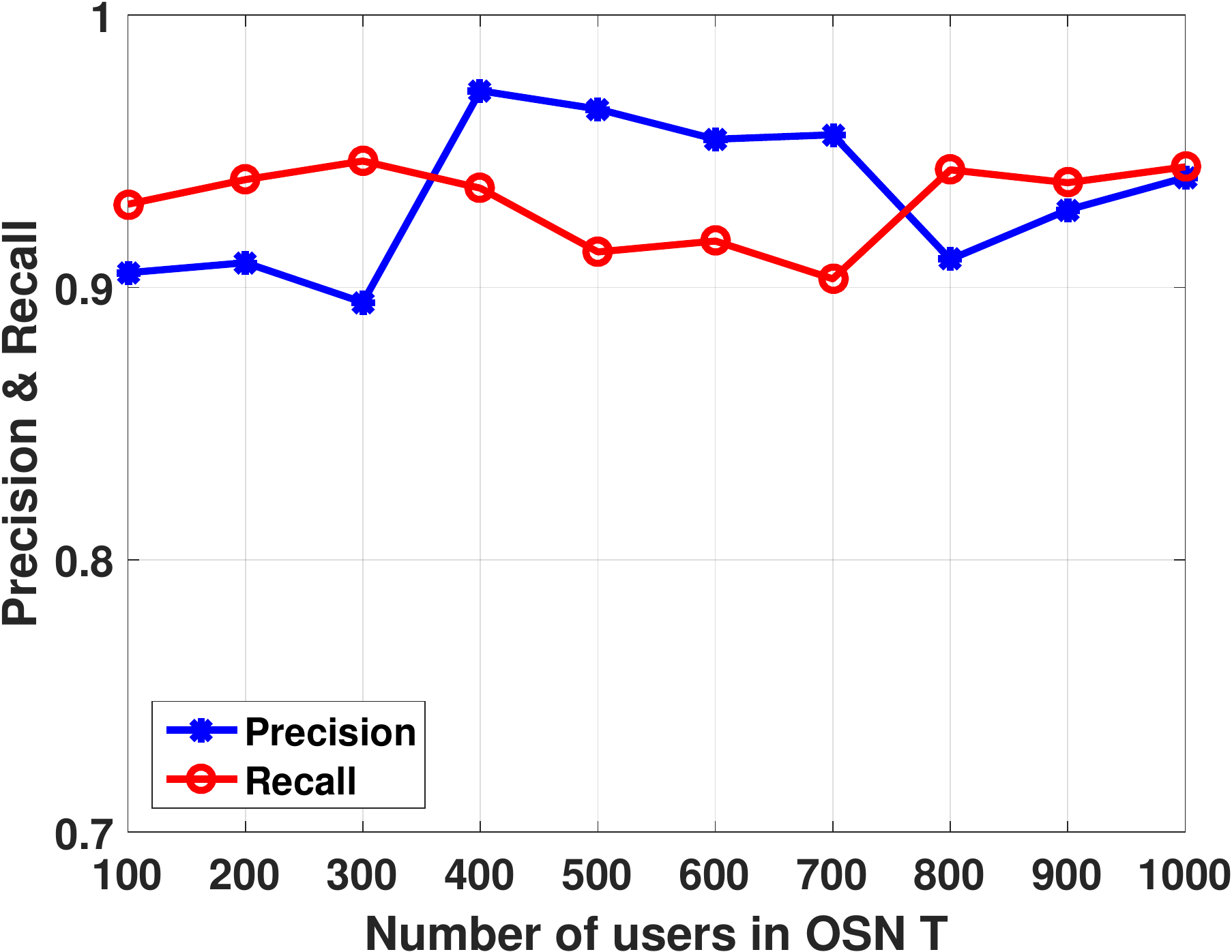}}
        \end{subfigure}\hfill
        \begin{subfigure}[$\mathrm{D2}$]{\includegraphics[scale=0.21]{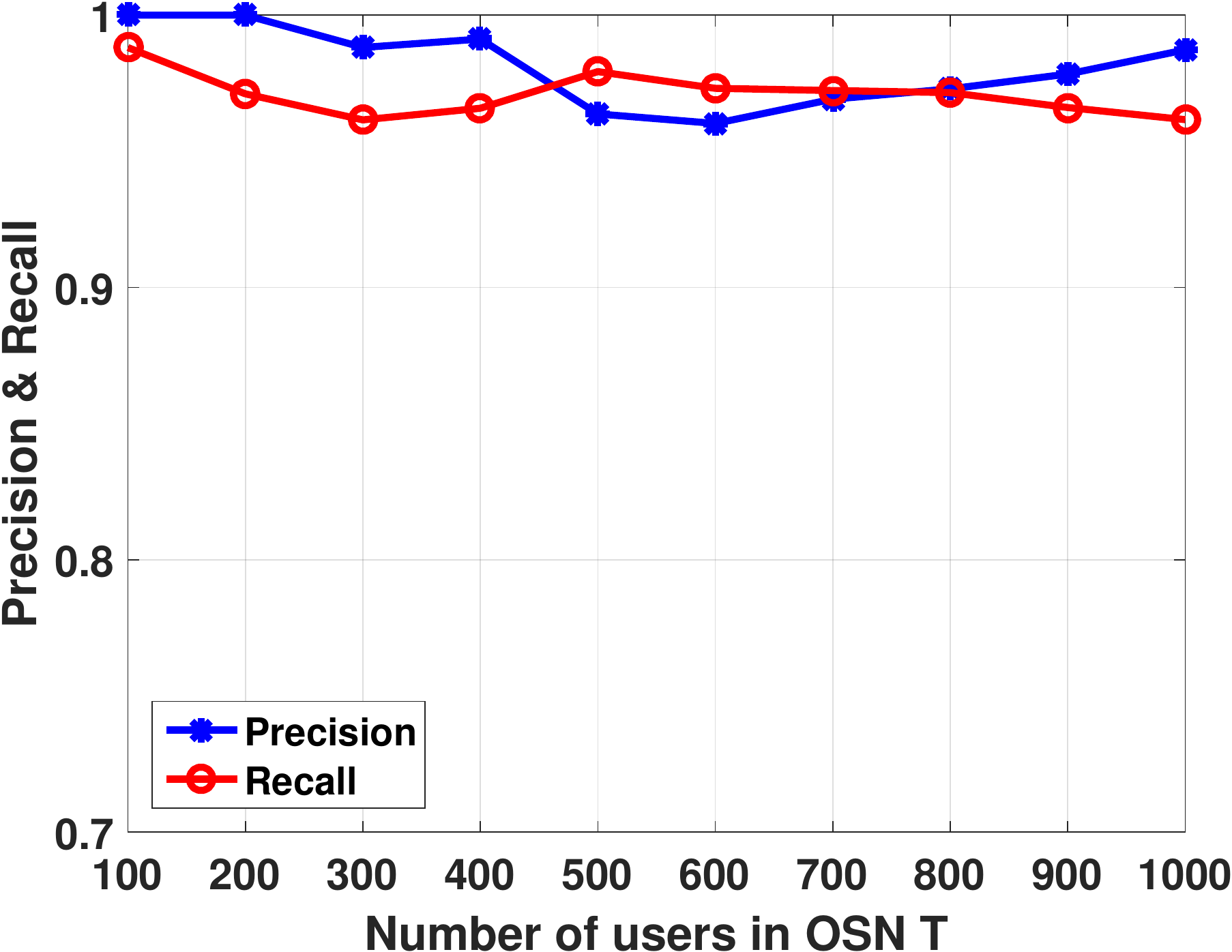}}
        \end{subfigure}\hfill
        \begin{subfigure}[$\mathrm{D3}$]{\includegraphics[scale=0.21]{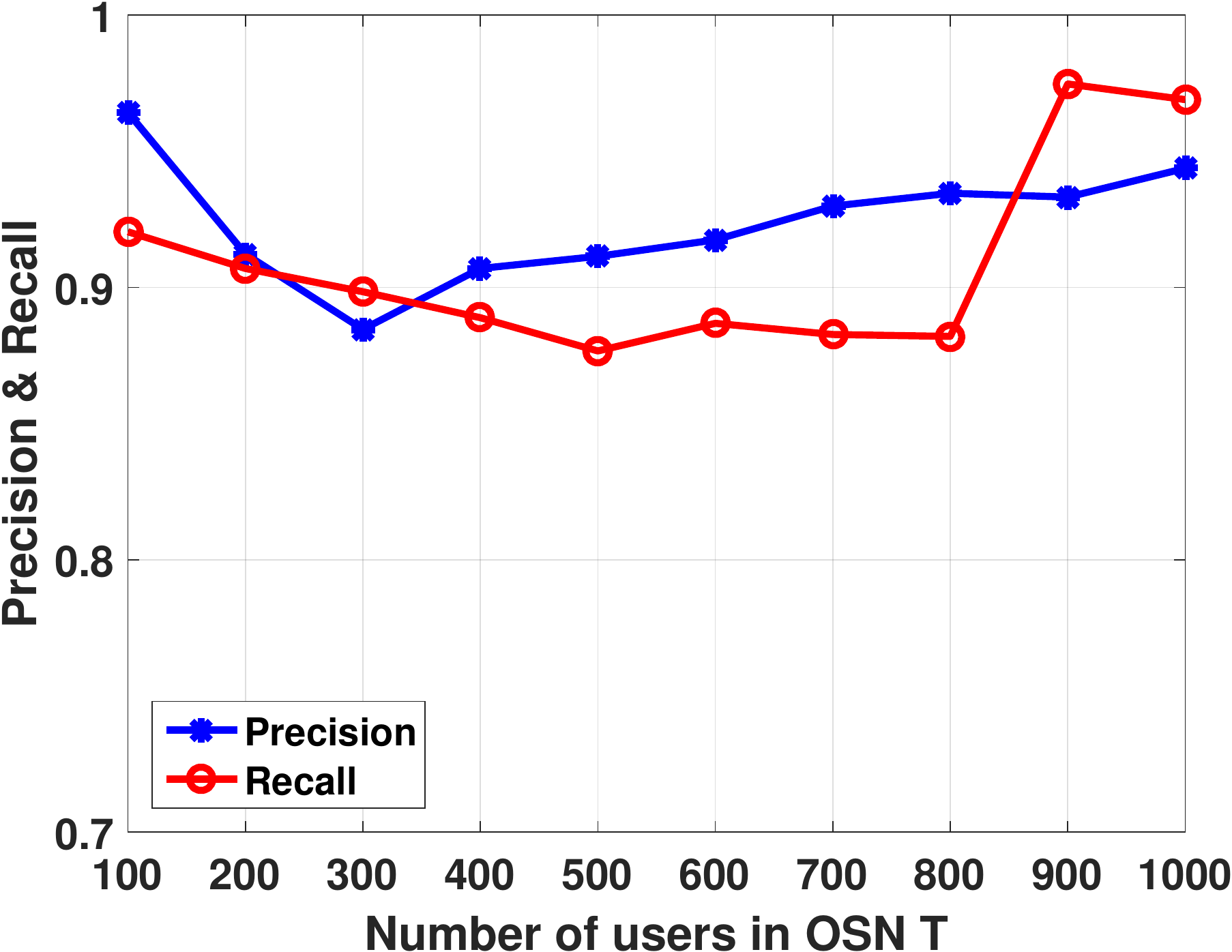}}
        \end{subfigure}\hfill
        \caption{The effect of target OSN's (OSN $T$) size to precision/recall when the size of auxiliary OSN (OSN $A$) is $1000$ in $\mathrm{D1}$, $\mathrm{D2}$, and $\mathrm{D3}$.} 
        \label{fig:fixed_osn_a}
\vspace{-10pt}
\end{figure}

We study the effect of the OSNs' size to precision and recall of the proposed algorithm. In Section~\ref{sec:BP_results}, we provided the results when the number of users in OSN $T$ is fixed. Here, we provide the results of the other two scenarios.
In Figure~\ref{fig:fixed_osn_a}, for each dataset, we show the precision/recall values of the BP-based algorithm when the number of users in the target OSN (OSN $T$) increases while the number of auxiliary users (i.e., users in OSN $A$) is fixed.
We set the number of users in OSN $A$ as $1000$ and increase the number of users in OSN $T$ from $100$ to $1000$ in steps of $100$.

\begin{figure}[t]
        \centering
        \begin{subfigure}[$\mathrm{D1}$]{\includegraphics[scale=0.21]{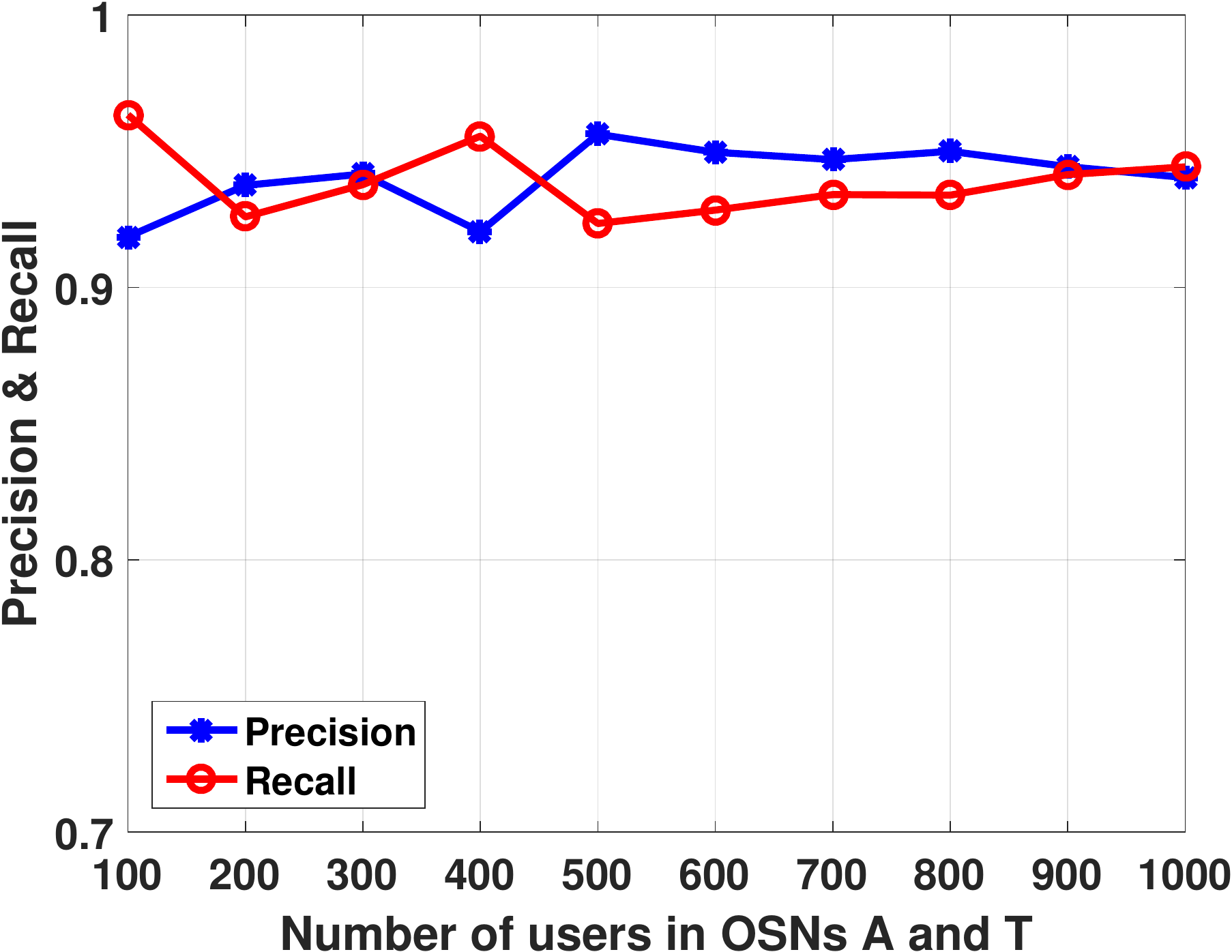}}
        \end{subfigure}\hfill
        \begin{subfigure}[$\mathrm{D2}$]{\includegraphics[scale=0.21]{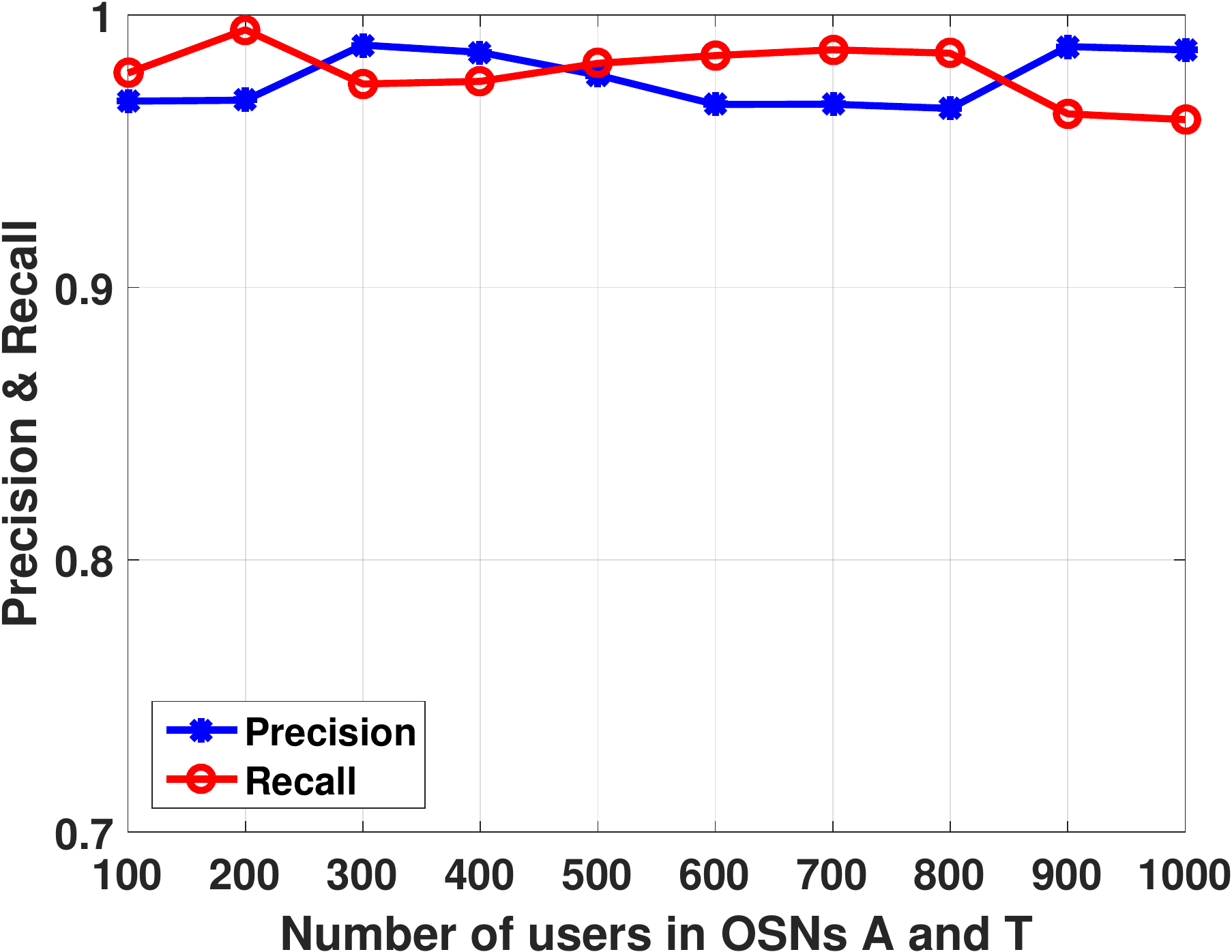}}
        \end{subfigure}\hfill
        \begin{subfigure}[$\mathrm{D3}$]{\includegraphics[scale=0.21]{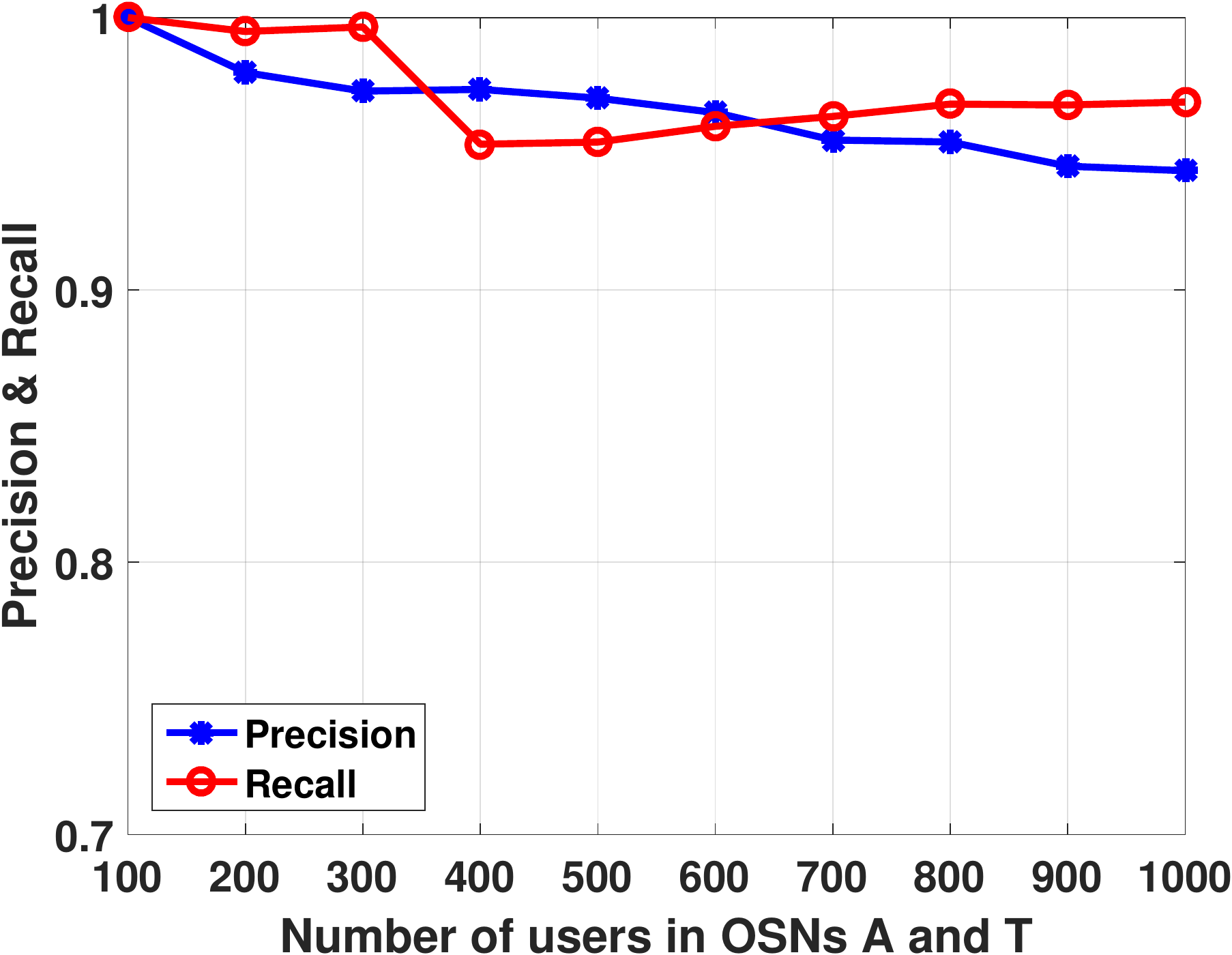}}
        \end{subfigure}\hfill
        \caption{The effect of auxiliary and target OSNs' (OSN $A$ and $T$) size to precision/recall in $\mathrm{D1}$, $\mathrm{D2}$, and $\mathrm{D3}$.} 
        \label{fig:both_osns}
\vspace{-10pt}
\end{figure}

In Figure~\ref{fig:both_osns}, for each dataset, we show the precision/recall values of the BP-based algorithm when the number of users in both OSNs (i.e., OSN $A$ and $T$) increases from $100$ to $1000$ in steps of $100$. In both scenarios, we observe that the precision/recall values of the proposed algorithm only slightly decrease with the increase of the number of users in the target OSN, or the increase of the number of users in both OSNs, which shows the scalability of our proposed algorithm. 

\vspace{70pt}
\begin{wrapfigure}{R}{0.4\textwidth}
	\centering
	\includegraphics[scale=0.21]{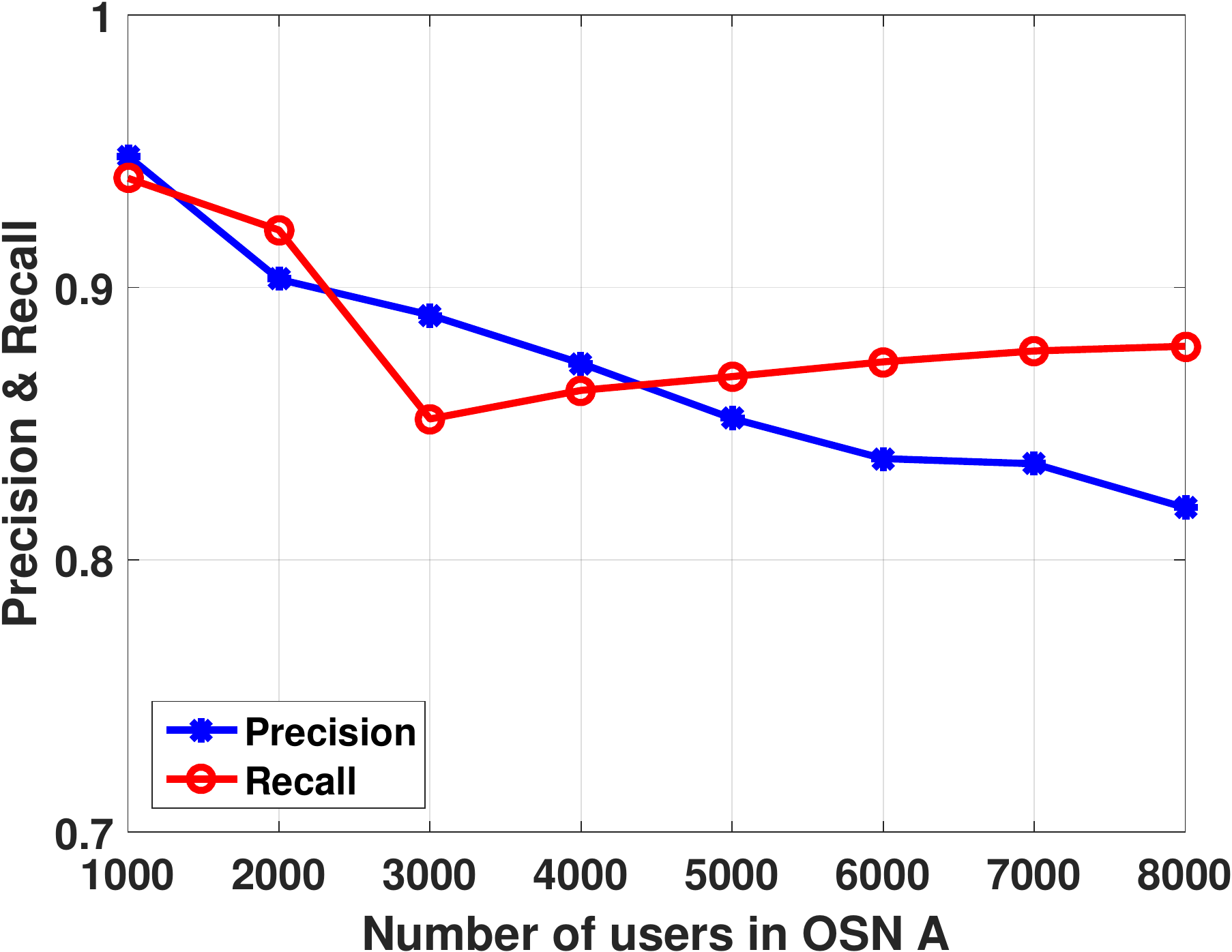}
	\caption{The effect of auxiliary OSN's (OSN $A$) size to precision/recall when the size of target OSN (OSN $T$) is $1000$ in $\mathrm{D3}$.}
	\label{fig:fixed_osn_t_large_dt}
	\vspace{-10pt}
\end{wrapfigure}
To further check the effect of the auxiliary OSN's size to precision and recall of the BP-based algorithm, we quantify the precision/recall values obtained by the proposed algorithm for larger scales in $\mathrm{D3}$. 
We fix the number of users in the target OSN (i.e., OSN $T$) to $1000$ while the number of users in the auxiliary OSN (i.e., OSN $A$) increases from $1000$ to $8000$ in steps of $1000$ (in Figure~\ref{fig:fixed_osn_t} the number of users in OSN $T$ was fixed to $100$ while the number of users in OSN $A$ was increasing from $100$ to $1000$). We show the results for $\mathrm{D3}$ in Figure~\ref{fig:fixed_osn_t_large_dt}. The precision/recall values slightly decrease with the increase of the number of users in OSN A, confirming the scalability of the proposed algorithm. Note that, in $\mathrm{D3}$, we only use the graph connectivity attribute for profile matching. We expect that the decrease in precision/recall values will be smaller when both the graphical structure and other attributes of the users are used to generate the model.

\end{document}